\documentclass[]{elsart}

\usepackage{graphicx}

\usepackage{mathptmx, courier, pifont}
\usepackage[scaled=0.92]{helvet}
\usepackage[T1]{fontenc}
\usepackage{textcomp}

\usepackage{bm}

\newcommand{\putfig}[6]{%
\begin{figure}\vspace*{#2}%
\begin{center}%
\includegraphics*[scale=#4]{#5}%
\end{center}%
\caption{#6}%
\vspace*{#3}%
\label{fig:#1}%
\end{figure}}

\newcommand{\tsub}[1]{_{\mbox{\scriptsize#1}}}
\newcommand{\tsup}[1]{^{\mbox{\scriptsize#1}}}

\newcommand{\quarterthin}{\kern 0.0417em}

\newcommand{\halfthin}{\kern 0.0834em}
\newcommand{\neghalfthin}{\kern -0.0834em}
\newcommand{\negquarterthin}{\kern - 0.0417em}

\newcommand{\diffelement}[1]{d\negquarterthin#1}
\newcommand{\deriv}[2]{\frac{\diffelement#1}{\diffelement#2}}

\newcommand\simgreater{\,\lower0.7ex\hbox{$\stackrel{>}{\sim}$}\,}
\newcommand\simless{\,\lower0.7ex\hbox{$\stackrel{<}{\sim}$}\,}

\newcommand{\tfrac}[2]{\mbox{$\small\frac{#1}{#2}$}}

\newcommand{\units}[1]{\mbox{\ #1}}

\newlength{\figdn}
\newcommand{\Fplus}[1]{F^+_{#1}}
\newcommand{\Fminus}[1]{F^-_{#1}}
\newcommand{\fplus}[1]{f^+_{#1}}
\newcommand{\fminus}[1]{f^-_{#1}}
\newcommand{\PEsource}[2]{f_{{#1}{\rightleftharpoons}{#2}}}
\newcommand{\eq}[1]{Eq.~(\ref{#1})}
\newcommand{\eqnoeq}[1]{(\ref{#1})}
\newcommand{\fig}[1]{Fig.~\ref{fig:#1}}

\newcommand{\isotope}[2]{\mbox{$^{#1}$#2}}

\begin{document}

\begin{frontmatter}

\title
{Algebraic Stabilization of Explicit Numerical Integration for Extremely Stiff
Reaction Networks}

\author{Mike Guidry}
\ead{guidry@utk.edu}

\address{Department of Physics and Astronomy, University of Tennessee, 
Knoxville, TN 37996-1200, USA}

\address{Physics Division, Oak Ridge National Laboratory, Oak Ridge, 
TN 37830, USA}

\address{Computer Science and Mathematics Division, Oak Ridge National 
Laboratory, Oak Ridge, TN 37830, USA}

\date{\today}

\begin{abstract}

In contrast to the prevailing view in the literature, it is shown that even
extremely stiff sets of ordinary differential equations may be solved
efficiently by explicit methods if limiting algebraic solutions are used to
stabilize the numerical integration. The stabilizing algebra differs essentially
for systems well-removed from equilibrium and those near equilibrium.  Explicit
asymptotic and quasi-steady-state methods that are appropriate when the system
is only weakly equilibrated are examined first. These methods are then extended
to the case of close approach to equilibrium  through a new implementation of
partial equilibrium approximations. Using stringent tests with astrophysical
thermonuclear networks, evidence is provided that these methods can deal with
the stiffest networks, even in the approach to equilibrium, with accuracy and
integration timestepping comparable to that of implicit methods. Because
explicit methods can execute a timestep faster and scale more favorably with
network size than implicit algorithms, our results suggest that
algebraically-stabilized explicit methods might enable integration of larger
reaction networks coupled to fluid dynamics than has been feasible previously
for a variety of disciplines.

\end{abstract}

\begin{keyword}
ordinary differential equations 
\sep 
reaction networks 
\sep 
stiffness
\sep
reactive flows
\sep
nucleosynthesis
\sep
combustion

\vspace{1ex}

\PACS 
02.60.Lj 
\sep
02.30.Jr 
\sep
82.33.Vx 
\sep
47.40 
\sep
26.30.-k 
\sep
95.30.Lz 
\sep
47.70.-n 
\sep
82.20.-w 
\sep
47.70.Pq 

\end{keyword}

\end{frontmatter}

\section{\label{intro} Introduction}

Problems from many disciplines require solving large coupled reaction networks.
Representative examples include  reaction networks in combustion chemistry
\cite{oran05}, geochemical cycling of elements \cite{magick}, and thermonuclear
reaction networks in astrophysics \cite{hix05,timmes}.  The differential
equations used to model these  networks usually exhibit stiffness, which
arises from multiple timescales in the problem that differ by many orders of
magnitude \cite{oran05,gear71,lamb91,press92}. Sufficiently complex physical
systems often involve important processes operating on widely-separated
timescales, so realistic problems tend to be at least moderately stiff. Some,
such as astrophysical thermonuclear networks, are extremely stiff, with 10--20
orders of magnitude between the fastest and slowest timescales in the problem.
Our concern here is with stiffness as a numerical issue,  but we remark that
stiffness can have important physical implications because complex processes
often function as they do precisely because of the coupling of very slow and
very fast scales within the same system. 

Books on numerical and computational methods routinely state
\cite{oran05,lamb91,press92} that stiff systems cannot be integrated efficiently
using explicit finite-difference methods because of stability issues: for an
explicit algorithm, the maximum stable timestep is set by the fastest
timescales, even if those timescales are peripheral to the main phenomena of
interest. The standard resolution of the stiffness problem uses implicit
integration, which is stable for stiff systems but entails substantial
computational overhead because it requires the inversion of matrices at each
integration step. Because of the matrix inversions, implicit algorithms  tend to
scale from quadratically to cubically with network size unless favorable matrix
structure can be exploited.  Thus, implicit methods can be expensive  for large
networks.

A simple but instructive example of stiffness is provided by the CNO cycle for
conversion of hydrogen to helium, which powers main-sequence stars more massive
than the Sun (Fig.~\ref{fig:cnoCycle}). In the CNO cycle the fastest rates under
characteristic stellar conditions are $\beta$-decays with half-lives $\sim 100$
seconds, but to track the complete evolution of main-sequence hydrogen burning
may require integration of the network for hydrogen burning over timescales as
large as billions of years ($\sim 10^{16}$ seconds). If one tries to implement
this integration using explicit forward differencing, the largest stable
integration timestep will be set by the fastest rates and will be of order
$10^2$ seconds.  Thus $10^{14}$ or more explicit integration steps could be
required to integrate the CNO cycle to hydrogen depletion. Conversely, typical
implicit integration schemes can take stable and accurate timesteps equal to
1-10\% of the local time over most of the integration range, and would  compute
the above numerical integration in a few hundred implicit steps. By virtue of
examples such as this, it is broadly accepted that explicit methods are not
viable for stiff networks. To quote the authoritative reference {\em Numerical
Recipes} \cite{press92}, ``For stiff problems we {\em must} use an implicit
method if we want to avoid having tiny stepsizes.''
\putfig
{cnoCycle}
{0pt}
{\figdn}
{0.90}
{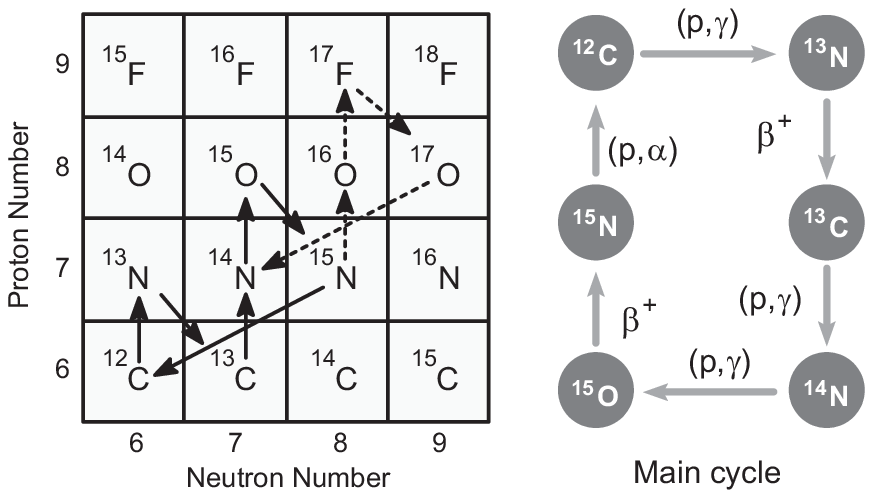}
{The CNO (carbon--nitrogen--oxygen) cycle. On the left side the main branch of
the cycle is illustrated with solid arrows and a side branch is illustrated with
dashed arrows. On the right side, the main branch of the CNO cycle is
illustrated in more detail.}

Our main interest lies in simulations where the reaction network is one part of
a broader problem.  Let us take as representative astrophysical thermonuclear
networks coupled to multidimensional hydrodynamics. The hydrodynamical evolution
controls network conditions (temperature, density, \ldots), and the network
influences the hydrodynamics through energy production and modification of
composition. Solution of large networks by the usual means is costly in this
context and even ambitious simulations  use only small networks, or replace the
network entirely by parameterization. Then a more realistic network is used in a
separate ``post-processing'' step, where fixed hydrodynamical profiles computed
in the original simulation specify the variation of temperature and density with
time. Such approximations are especially at issue for problems
like Type Ia supernovae, which are 3D asymmetric explosions
powered by a complex reaction network releasing energy greater than that of a
large galaxy on a timescale of order 1 s.

Astrophysical reaction networks have been used to illustrate, but problems in
various fields exhibit a similar complexity.  For example, in astrochemical
kinetics large chemical evolution networks must be modeled in dynamical
environments such as contracting molecular clouds, or in combustion chemistry
the burning networks are strongly coupled to dynamical simulations of the air
and fuel mixture. Realistic networks in all such applications may be quite
large. Modeling combustion of larger hydrocarbon molecules or soot formation can
require hundreds to thousands of reacting species with up to 10,000 reactions
\cite{oran05}, and realistic networks for supernova explosions imply hundreds to
thousands of nuclear isotopes with tens of thousands of reaction couplings
\cite{hix05}. In all such cases  current techniques do not permit the coupling
of realistic reaction networks to the full dynamics and highly-schematic
reaction networks are used in even the most realistic contemporary simulations.

\section{\label{sh:varieties} Stiffness under Equilibrium and Non-Equilibrium
Conditions}

The pessimism  engendered by the Introduction notwithstanding, it would be
highly desirable to integrate large, complex networks by explicit means because
explicit algorithms are simple and economical, and scale favorably with network
size. In principle, this could be accomplished by identifying conditions under
which some equations in the network have an approximate analytical solution and
using that information to remove algebraically the stiffest components from the
general numerical solution, thereby replacing the original network with an
approximation that permits larger stable explicit-integration timesteps. To that
end, the first task is to understand clearly the nature of the stiffness that we
wish to remove  from the equations.

\subsection{Varieties of Stiffness}

There are several fundamentally different sources of stiffness instability in
reaction networks that are often not clearly distinguished in the literature.
Pedagogically, discussions often emphasize an instability associated with small
populations becoming negative because of an overly-ambitious numerical
integration step, thus converting stable decaying-exponential terms into
unstable growing exponentials. However, there are other instabilities that can
occur, even when no population variables become negative in an integration
step. 
In these types of instabilities one ends up having to take the difference of
large numbers to obtain a result very near zero.  The numerical errors that
ensue in a standard explicit approach can then accumulate rapidly and
destabilize the network, even before any populations become negative. This is
still a stiffness instability because the problem results ultimately from a
numerical integration attempting to cope with highly-disparate timescales, but
the origin of these timescales is different from that discussed above. In this
case the disparate timescales are the fast reactions driving the system to
equilibrium contrasted with the slow timescale associated with equilibrium
itself (which tends to infinity). To illustrate, consider the form of the
equations that we desire to solve:
 \begin{eqnarray}
\deriv{y_i}{t} &=&
\Fplus{i} - \Fminus{i} 
\nonumber
\\
&=& (\fplus 1 + \fplus 2 + \ldots)_i  - (\fminus 1 + \fminus 2 + \ldots)_i
\nonumber
\\
&=& (\fplus 1 - \fminus 1)_i + (\fplus 2 - \fminus 2)_i + \ldots 
 = \sum_j (\fplus j - \fminus j)_i,
\label{equilDecomposition}
\end{eqnarray}
where the $y_i (i=1 \dots N)$ describe the dependent variables (species
abundances for our examples), $t$ is the independent variable (time in our
examples), the fluxes between species $i$ and $j$ are denoted by
$(f^{\pm}_j)_i$, and the sum for each variable $i$ is over all variables $j$
coupled to $i$ by a non-zero flux $(f^{\pm}_{j})_i$. For an $N$-species network
there will be $N$ such equations in the populations $y_i$, with these equations
generally coupled to each other because of the dependence of the fluxes on the
different $y_j$.  

In \eq{equilDecomposition} several different ways to group the terms on the
right side have been indicated, with the first line representing a decomposition
into total flux in and out of species $i$ and the third line separating the
total flux into in and out contributions from individual reactions.  The utility
of these alternative decompositions will be elaborated further below.

\subsection{\label{ss:approachEquil} The Approach to Equilibrium}

In this discussion we shall employ the term ``equilibration'' broadly to mean
that evolution of the network is being  influenced strongly by nearly-canceling
terms on the right sides of the differential equations
\eqnoeq{equilDecomposition}. Two qualitatively different equilibrium conditions
may be distinguished that require different algebraic approaches to
stabilization.

\subsubsection{\label{ss:macroscopic} Macroscopic Equilibration}

One class of equilibrium conditions results if $\Fplus i - \Fminus i \rightarrow
0$ (asymptotic) or $\Fplus i - \Fminus i \rightarrow$ constant (steady-state).
Let us refer to these conditions as a {\em macroscopic equilibration}, since
they are statements about entire differential equations in
\eq{equilDecomposition}. We shall introduce asymptotic and quasi-steady-state
approximations exploiting these conditions that remove whole differential
equations from the numerical integration for a network timestep, replacing them
with algebraic solutions. Such approximations still integrate the full original
set of equations, but they reduce the number of equations integrated {\em
numerically by forward difference}.  This helps with stiffness because
integration of some or all of the stiffest equations is replaced by a stable
analytical solution, and any equations that remain to be integrated numerically
tend to have smaller disparities in timescales and thus less stiffness.

\subsubsection{\label{ss:microscopic} Microscopic Equilibration}

In \eq{equilDecomposition}, $\Fplus i$ and $\Fminus i$ for a  species $i$ each
consist of various terms depending on the other populations in the network.
Groups of individual terms on the right side of \eq{equilDecomposition} may come
into equilibrium (so that the sum of their fluxes tends to zero), even if the
macroscopic conditions for equilibration for the entire equation are not
satisfied. We shall term this {\em microscopic equilibration}. Then we may
consider an algebraic approximation that removes groups of such terms from the
numerical integration by replacing their sum of fluxes with zero. This does not
reduce the number of equations integrated numerically, but can reduce their
stiffness by removing terms with fast rates, thereby reducing the disparity
between the fastest and slowest timescales in the system. Such considerations
will be the basis of the partial equilibrium methods to be discussed in
\S\ref{partialEq}.

\section{Reaction Networks in the Context of Larger Systems}

Assume the coupling of reaction networks to a larger system to be done using
operator splitting, where the larger system (the hydrodynamical solver in our
examples) is evolved for a timestep holding the parameters computed from the
network in the previous step constant, and then the network is evolved over the
time corresponding to the hydrodynamical timestep while holding the new
variables calculated in the hydrodynamical timestep constant.  This places two 
strong constraints on methods:

\begin{enumerate}

 \item 
The network must be advanced with new initial conditions for each hydrodynamical
step.  Thus, algorithms must initialize simply and quickly. 

\item
Integration of the network must not require time substantially larger than
that for the corresponding hydrodynamical timestep.  

\end{enumerate}
Let us elaborate further on this second point. Taking the multidimensional,
adaptive-mesh, explicit hydrodynamical code Flash \cite{fry00} applied to Type
Ia supernova explosions as representative,  we estimate in Ref.\ \cite{guidAsy}
that a simulation with realistic networks can be done in a tractable amount of
time if the algorithm can take  timesteps $\Delta t$ over the full
hydrodynamical integration range that average at least $(0.01 - 0.001) \times
t$, where $t$ is the elapsed time in the integration. Timesteps of this size are
possible with implicit and semi-implicit algorithms, but those methods are
inefficient at computing each timestep in large networks; explicit methods can
compute a timestep efficiently, but timesteps this large are unthinkable with a
normal explicit algorithm because they are typically unstable. In this paper we
shall demonstrate explicit integration methods that realize such large
integration timesteps in a variety of extremely stiff examples. This alters
essentially the discussion of whether explicit methods, with their faster
computation of timesteps and more favorable scaling with network size, are
practical for large, stiff networks. The general approximations to be
discussed have been implemented before \cite{oran05,youn77,mott00,mott99}, but
we
shall find that our implementations are much more successful than previous
applications to extremely stiff networks; accordingly we shall reach rather
different conclusions about these methods than those of earlier publications.

\section{\label{algebraic} Algebraic Stabilization of Solutions Far from
Equilibrium}

Let us  address first explicit approximations that are appropriate if the system
is far from microscopic equilibrium. (Methods to determine whether this
condition is satisfied will be discussed in \S\ref{partialEq}.)

\subsection{\label{ss:asymptoticSimple} Explicit Asymptotic Approximations}

The differential equations to be solved take the  form given by
\eq{equilDecomposition}. Generally, $\Fplus i$ and $\Fminus i$ for a given
species $i$ each consist of a number of terms depending on the other populations
in the network. For the networks that we shall consider the depletion flux for
the population $y_i$ will be proportional to $y_i$,
\begin{equation}
    \Fminus i = (k_1^i + k_2^i + \ldots + k_m^i)y_i \equiv k^iy_i,
\label{eq1.2}
\end{equation}
where the $k_j^i$ are rate parameters (in units of time$^{-1}$) for each of the
$m$ processes that can deplete $y_i$, which may  depend on the populations $y_j$
and on system variables such as temperature and density. The characteristic
timescales $\tau^i_j = 1/k^i_j$ will vary over many orders of magnitude in the
systems of interest, meaning that these equations are very stiff. From
\eq{eq1.2} the effective total depletion rate $k^i$ for  $y_i$ at a given time,
and a corresponding timescale $\tau^i$, may be defined as
\begin{equation}
    k^i \equiv  \frac{\Fminus i}{y_i}
\qquad
\tau^i = \frac{1}{k^i},
\label{eq1.3}
\end{equation}
permitting \eq{equilDecomposition} to be written as
\begin{equation}
    y_i = \frac{1}{k^i} \left( \Fplus i - \deriv{y_i}{t} \right).
\label{eq1.4}
\end{equation}
Thus, in a finite-difference approximation at timestep $t_n$
\begin{equation}
    y_i(t_n) = \frac{\Fplus i (t_n)}{k^i(t_n)} - \frac{1}{k^i(t_n)} \left.
\deriv{y_i}{t}
    \right|_{t=t_n}.
\label{eq1.5}
\end{equation}
The {\em asymptotic limit} for the species $i$ corresponds to the condition
$\Fplus i \simeq \Fminus i$, implying from \eq{equilDecomposition} that
$\diffelement y_i /\diffelement t \simeq 0$.  In this limit \eq{eq1.5} gives a
first approximation $ y^{(1)}_i (t_n)$ and local error $E_n^{(1)}$,
respectively, for $y_i(t_n)$,
\begin{equation}
 y^{(1)}_i (t_n) = \frac{\Fplus i(t_n)}{ k^i(t_n)}
\qquad
E_n^{(1)} \equiv y(t_n) - y^{(1)}(t_n) = -\frac{1}{k(t_n)}
\deriv yt(t_n) .
\label{firstApprox}
\end{equation}
For small $\diffelement y_i/\diffelement t$ a correction term may be obtained by
writing the derivative term in \eq{eq1.5} as
\begin{equation}
 \deriv{y}{t}(t_n) = \frac{1}{\Delta t}
    \left( y_i(t_n) - y_{i}(t_{n-1}) \right)
+\frac{1}{\Delta t} \left( E_n^{(1)} - E_{n-1}^{(1)}\right) +  O(\Delta t) ,
\label{derivative}
\end{equation}
where $O(x)$ denotes terms of order $x$. Substitution in  \eq{eq1.5} then gives
$$
y^{(2)}_n = \frac{F^+_n}{k_n} -\frac{1}{k(t_n) \Delta t} \left( y(t_n) -
y(t_{n-1}
\right) 
- \frac{1}{k(t_n) \Delta t} \left( -E_n^{(1)} + E_{n-1}^{(1)}\right),
$$
and setting $y(t_n) = y^{(2)}_n$ and solving for $y^{(2)}_n$ gives
\begin{equation}
y_n^{(2)} = \frac{1}{1+k_n\Delta t} \left(y_{n-1} + F^+_n \Delta t \right),
\label{asySophia}
\end{equation}
where a term of order $(\Delta t) ^2/(1+k(t_n)\Delta t)$ has been discarded.
This approximation is expected to be valid for large $k\Delta t$. Another
approach is to use a predictor--corrector scheme within such an asymptotic
approximation \cite{oran05}. However, it has been shown  \cite{guidAsy} that
these asymptotic approximations give rather similar results for the networks
that we shall test, so the simple formula \eqnoeq{asySophia}  will be adequate
for the present discussion. The mathematical and numerical properties of
asymptotic approximations have been explored previously in Refs.\
\cite{oran05,youn77,mott00,mott99}.

To implement an asymptotic algorithm we define a critical value $\kappa$ of
$k\Delta t$ and at each timestep cycle through all populations and compute the
product $k^i\Delta t$ for each species $i$ using \eq{eq1.3} and the proposed
timestep $\Delta t$.   Then, for each species $i$
\begin{enumerate}
\item
If $k^i\Delta t < \kappa$, the population is updated numerically by the 
explicit Euler method.
\item
If $k\Delta t \ge \kappa$, the population is updated
algebraically using \eq{asySophia}.
\end{enumerate}
Formally explicit integration is expected to be stable if $k^i\Delta t <1$ and
potentially unstable if $k^i\Delta t \ge 1$ (see the discussions in Refs.\
\cite{oran05} and \cite{guidAsy}). Therefore, $\kappa = 1$ has been chosen for
all examples presented here.

\subsection{\label{sh:steadystate} Quasi-Steady-State Approximations}

 An alternative explicit algebraic solution to the coupled differential
equations is possible using the quasi-steady-state (QSS) approximations
developed by Mott and collaborators \cite{mott00,mott99}, which drew on earlier
work in Refs.\ \cite{verw94,verw95,jay97}. Following Mott et al
\cite{mott00,mott99}, first notice that \eq{equilDecomposition}, expressed in
the form
$
dy/dt = F^+ (t) - k(t) y(t)
$ 
using \eq{eq1.2} and with indices dropped for notational convenience, has the
general solution
\begin{equation}
 y(t) = y_0 e^{-kt} + \frac{F^+}{k}(1-e^{-kt}),
\label{qssSolution}
\end{equation}
provided that $k$ and $F^+$ are constant. The QSS method then uses this equation
as the basis of a predictor--corrector algorithm in which a prediction is made
using only initial values and then a corrector is applied that uses a
combination of initial values and values computed using the predictor solution.
In terms of a parameter $\alpha(r)$ defined by
\begin{equation}
\alpha(r) = \frac{160 r^3 + 60 r^2 + 11r +1}{360r^3 + 60r^2 + 12r +1},
\label{qss1.1}
\end{equation}
where $r \equiv 1/k\Delta t$, we adopt a predictor $y\tsup p$ and 
corrector $y\tsup c$ given by \cite{mott00,mott99}
\begin{equation}
 y\tsup p = y^0 + \frac{\Delta t (F_0^+ - F_0^-)}{1 + \alpha^0 k^0 \Delta t}
\qquad
y\tsup c = y^0 + \frac{\tilde F^+ - \bar k y^0}{1+\bar\alpha \bar k \Delta t} ,
\label{qss1.2}
\end{equation}
where $\alpha^0$ is evaluated from \eq{qss1.1} with $r = 1/k^0 \Delta t$,
an average rate parameter is defined by 
$\bar k = \tfrac12 (k^0 + k\tsup p)$,
$\bar\alpha$ is specified by \eq{qss1.1} with $r = 1/\bar k\Delta t$, and
$$
 \tilde F^+ = \bar\alpha F_{\scriptstyle\rm p}^+ + (1-\bar\alpha)F_0^+.
$$
The corrector can be iterated if desired by using $y\tsup c$ from one iteration
step as the $y\tsup p$ for the next iteration step. We implement an explicit QSS
algorithm based on the predictor--corrector \eqnoeq{qss1.2} essentially in
parallel with that described above for the asymptotic method, except that in
applying the QSS algorithm all equations are treated in QSS approximation,
rather than dividing the equations into a set treated by explicit forward
difference and a set treated analytically \cite{guidQSS}.

\section{Algebraic Stabilization in the Approach to Equilibrium}
\protect\label{partialEq}

We shall find that the asymptotic and quasi-steady-state methods described in
preceding sections work well for macroscopic equilibration, but are highly
inefficient for microscopic equilibration.  Thus, these methods must be
augmented by a means to remove stiffness associated with the approach to
(microscopic) equilibrium if they are to be applicable to a broad range of
problems. In this section approximations to stabilize explicit integration in
the presence of microscopic equilibration are developed.  This development draws
heavily on the partial equilibrium work of David Mott \cite{mott99} as a
starting point, but we shall extend these methods and find much more favorable
results for extremely stiff networks than those obtained in the pioneering work
of Mott and collaborators \cite{guidPE}.

Partial equilibrium (PE) methods examine source terms $\fplus i$ and $\fminus i$
for individual reaction pairs in the network---not the composite fluxes $\Fplus
i$ and $\Fminus i$ that are the basis for asymptotic and QSS
approximations---for approach to equilibrium. When a fast reaction pair nears
equilibrium its source terms are removed from the direct numerical integration
in favor of an equilibrium algebraic constraint. Reactions not in equilibrium
still contribute to the fluxes for the numerical integrator, but once fast
reactions are decoupled from the numerical integration the remaining system
typically becomes (much) less stiff. Consider a representative 2-body reaction
and its source term $\PEsource{ab}{cd}$,
\begin{equation}
 a + b \rightleftharpoons c + d \qquad
\PEsource{ab}{cd} = \pm (k\tsub f y_a y_b - k\tsub r y_c y_d) ,
\label{partial1.1}
\end{equation}
where the $y_i$ denote population variables for the species $i$ and the $k$s are
rate parameters for forward ($k\tsub f$) and reverse ($k\tsub r$) reactions.
Considered in isolation, the reaction pair of \eq{partial1.1} may be deemed to
be in equilibrium if $\PEsource{ab}{cd} = 0$.  It is useful to introduce the
idea of {\em partial equilibrium (PE),} where at a given time some reaction
pairs have $f=0$ and some have $f \ne 0$. The evolution of the system is then
determined primarily by those reactions for which $f \ne 0$, but since the
system is coupled the $f\ne 0$ reactions will perturb the $f=0$ reaction pairs
so that for those reactions near equilibrium $f = 0 \rightarrow f \sim 0$.  Let
us now introduce a set of definitions and concepts that will allow us both to
quantify and use to our advantage the partial equilibrium condition $f \sim 0$.

\subsection{Conserved Scalars and Progress Variables}
\protect\label{progvar}

The reaction pair $a+b \rightleftharpoons c+d$ appears at first glance to have
four characteristic timescales associated with the rate of change for the four
populations $y_a$, $y_b$, $y_c$, and $y_d$, respectively. However, it is clear
that the following three constraints apply to this reaction
\begin{equation}
 y_a-y_b = c_1 \qquad y_a + y_c = c_2 \qquad y_a + y_d = c_3,
\label{2body1.1}
\end{equation}
where the constants $c_i$ may be evaluated by substituting the initial
abundances into \eq{2body1.1}, 
$$
c_1 = y_a^0 - y_b^0
\qquad
c_2 = y_a^0 + y_c^0
\qquad
c_3 = y_a^0 + y_d^0.
$$ 
Losing one $a$ in the reaction $a+b \rightleftharpoons c+d$ requires the
simultaneous loss of one $b$, so their difference must be constant, as implied
by the first of Eqs.\ \eqnoeq{2body1.1}, and every loss of one $a$ produces one
$c$ and one $d$, implying the second and third of Eqs.\ \eqnoeq{2body1.1}. The
left sides in the equations \eqnoeq{2body1.1} are examples of  {\em conserved
scalars} \cite{mott99}, which are constant by virtue of the structure of the
equations, not by any particular dynamical assumptions. The differential
equation describing the evolution of $y_a$ is
\begin{equation}
 \deriv{y_a}{t} = -k\tsub f y_a y_b + k\tsub r y_c y_d
= a y_a^2 + b y_a + c,
\label{2body1.3}
\end{equation}
where \eq{2body1.1} has been used and
$$
a \equiv k\tsub r - k\tsub f \qquad b \equiv -k\tsub r(c_2+c_3) + k\tsub f c_1
\qquad c \equiv k\tsub r c_2 c_3.
$$
We shall demonstrate below that the approach to equilibrium for any 2-body
reaction pair can be described, and the approach to equilibrium for any 3-body
reaction pair can be approximated, by a differential equation of this form. In
terms of the quantity
\begin{equation}
 q \equiv 4ac-b^2 ,
\label{2body1.3a}
\end{equation}
the solutions to \eq{2body1.3} of interest in the present context correspond to
$a\ne0$ and $q < 0$, and take the form \cite{mott99,feg11b}
\begin{equation}
  y_a(t) = -\frac{1}{2a}\left(
b + \sqrt{-q} \,
\frac{1+\phi \exp (-\sqrt{-q} \,t)} {1-\phi \exp (-\sqrt{-q} \, t)}
\right) 
\quad
\phi \equiv \frac{2a y_0 + b + \sqrt{-q}} {2a y_0 + b - \sqrt{-q}}.
\label{2body1.4}
\end{equation}
The equilibrium solution then corresponds to the limit $t \rightarrow \infty$
of \eq{2body1.4},
\begin{equation}
 \bar y_a \equiv y^{{\rm\scriptstyle eq}}_a = -\frac{1}{2a} (b + \sqrt{-q}).
\label{2body1.6}
\end{equation}
Once $y_a$ has been determined the constraints \eqnoeq{2body1.1} may be used to
determine the other abundances.  For example, at equilibrium
$$
\bar y_b(t) = \bar y_a(t) -c_1
\qquad
\bar y_c(t) = c_2 -\bar y_a(t)
\qquad
\bar y_d(t) = c_3 - \bar y_a(t).
$$
It is often convenient to introduce a variable $\lambda$ that is the difference
between the values of the $y_i$ at the beginning of the timestep and their
current values
\begin{eqnarray}
 y_a = y_a^0 - \lambda
\qquad y_b = y_b^0 - \lambda\qquad
y_c = y_c^0 + \lambda \qquad
y_d = y_d^0 + \lambda.
\label{progress1.5}
\end{eqnarray}
The new variable $\lambda$ is termed a {\em progress variable} for the reaction
characterized by $\PEsource{ab}{cd}$ and satisfies
\begin{equation}
 \deriv{\lambda}{t} = \PEsource{ab}{cd} \qquad \lambda_0
\equiv \lambda(t=0) = 0.
\label{progress1.4}
\end{equation}
Thus the approach of  $a+b \rightleftharpoons c+d$ to equilibrium is controlled
by a single differential equation \eqnoeq{2body1.3} that can be expressed in
terms of either a single one of the abundances $y_i$, or the progress variable
$\lambda$. The general solution of this equation is of the form given by
\eq{2body1.4}, with the  time dependence residing dominantly in the
exponentials. Therefore, the rate at which $a+b \rightleftharpoons c+d$ evolves
toward the equilibrium solution \eqnoeq{2body1.6} is governed by a {\em single
timescale}
\begin{equation}
 \tau = \frac{1}{\sqrt{-q}},
\label{2body1.8}
\end{equation}
which is illustrated in \fig{equilTimescaleComposite}.
 \putfig
     {equilTimescaleComposite}   
     {0pt}
     {\figdn}
     {0.73}
     {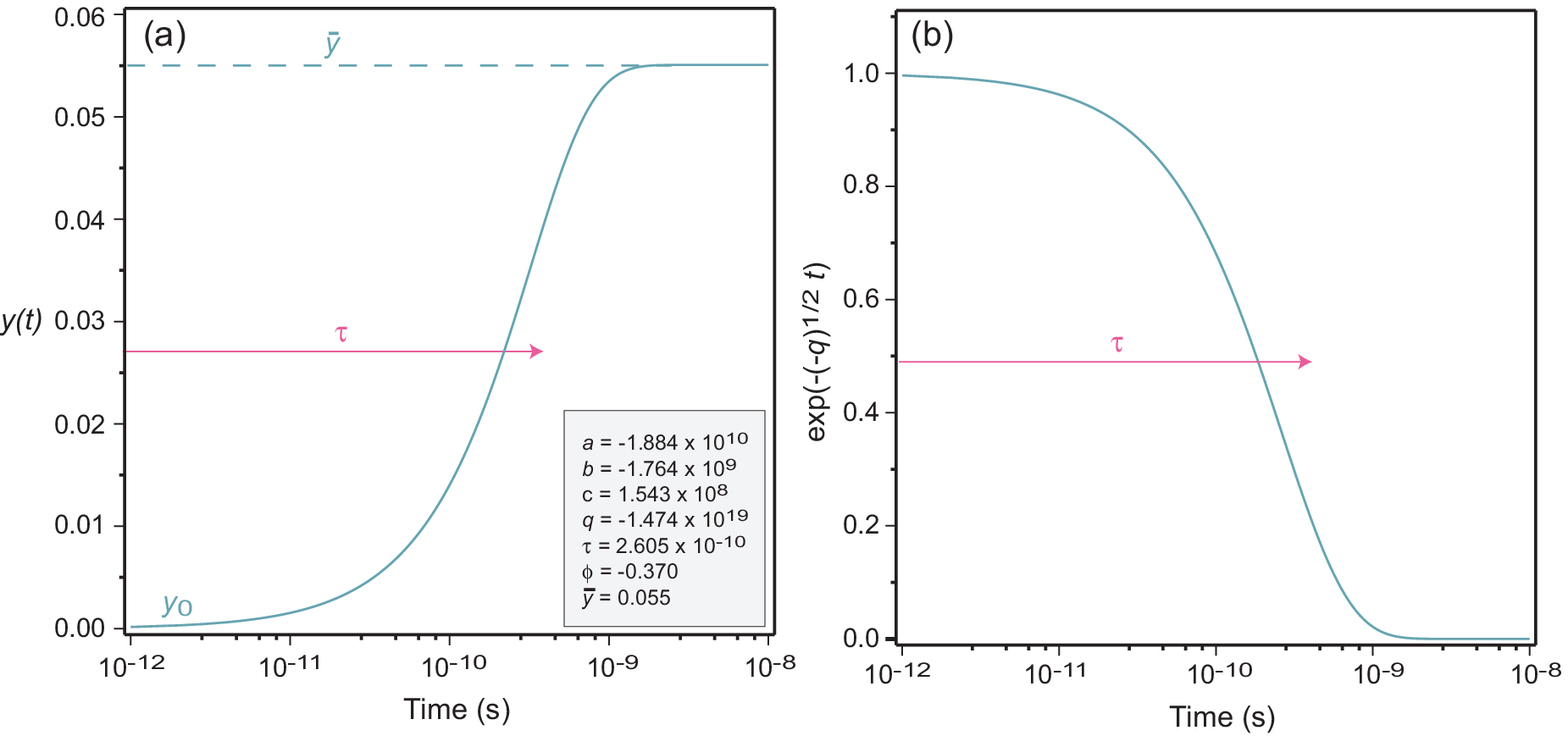}
{(a)~Time evolution of the solution \eqnoeq{2body1.4} assuming constant $a$,
$b$, and $c$. The characteristic timescale for approach to equilibrium
\eqnoeq{2body1.8} is labeled $\tau$ and the equilibrium value of $y(t)$ defined
by \eq{2body1.6} is denoted by $\bar y$. To illustrate we have assumed the
initial value $y_0 = 0$. For times considerably larger than $\tau$ the general
solution \eqnoeq{2body1.4} saturates at the equilibrium solution
\eqnoeq{2body1.6}. (b)~Behavior of the exponential factor in \eq{2body1.4}.}
Whether a reaction pair is near equilibrium at time $t$ may then be determined
by requiring that
\begin{equation}
 \frac{| y_i(t) - \bar y_i |}{\bar y_i}
< \epsilon_i
\label{2body1.8b}
\end{equation}
for each of the species $i$ that is involved in the reaction pair, where
$y_i(t)$ is the actual
abundance, $\bar y_i$ is the equilibrium abundance determined from
\eq{2body1.6},
and $\epsilon_i$ is a user-specified tolerance that could depend on $i$ but
will be taken to be the same for all species  in the examples discussed to
be here.
Alternatively, the equilibrium timescale \eqnoeq{2body1.8} may be compared  with
the current numerical timestep to determine whether a reaction is near
equilibrium: if $\tau$ is much smaller than the timestep, we may expect that
equilibrium can be established and maintained in successive timesteps, even if
it is being continually disturbed by other non-equilibrated processes.

\subsection{Reaction Vectors}
\protect\label{reactionvectors}

A partial equilibrium approximation in a large network could require that
thousands of reactions be examined for their equilibrium status at each
timestep.   Let us introduce a formalism, adapted from the work of Mott
\cite{mott99}, that allows examination of partial equilibrium criteria in a
particularly efficient way by exploiting the analogy of a reaction network to a
linear vector space. This will have two large advantages:  (1)~It will provide
us with some well-established mathematical tools.  (2)~The abstraction of
reaction network as linear vector space permits formulation of a partial
equilibrium algorithm that is not strongly tied to the details of a particular
problem, thus aiding portability within and across disciplines.

We begin by expressing the concentration variables for the $n$ species $A_i$ in
a network as components of a composition vector
\begin{equation}
  \bm y = (y_1, y_2, y_3, \ldots y_n),
\label{vector1.1}
\end{equation}
which lies in an $n$-dimensional vector space $\Phi$.  The components $y_i$ are
proportional to number densities for the species labeled by $A_i$, so a specific
vector in this space defines a particular composition. Any reaction in the
network can then be written in the form
$\sum_{i=1}^n a_i A_i \rightleftharpoons \sum_{i=1}^n b_i A_i$
for some sets of coefficients $\{a_i\}$ and $\{b_i\}$. 
The coefficients on the two sides of a reaction may be used to define a vector
$\bm r \in \Phi$ with components
\begin{equation} 
\bm r \equiv (b_1-a_1, b_2-a_2, \ldots, b_n-a_n),
\label{reactionVector}
\end{equation}
that specifies how the composition may change because of the reaction. For
example, consider the main part of the CNO cycle illustrated on the right side
of Fig.~\ref{fig:cnoCycle}. Choosing a basis $\{ p, \alpha, \isotope{12}C,
\isotope{13}N, \isotope{13}C, \isotope{14}N, \isotope{15}O, \isotope{15}N \},$
the
reaction $p + \isotope{15}{N} \rightarrow \alpha + \isotope{12}C$ then has a
reaction vector $\bm r$ with components
$
(-1, 1, 1, 0, 0, 0, 0, -1).
$

For a network with three or fewer species the corresponding linear vector space 
can be displayed geometrically.  For example, Fig.~\ref{fig:reactionVectors3}
illustrates a network containing the isotopes $\{\isotope{4}{He},
\isotope{8}{Be}, \isotope{12}{C}\}$ and  four reaction vectors corresponding to
the reaction pairs $2\alpha \rightleftharpoons \isotope{8}{Be}$ and $\alpha +
\isotope{8}{Be} \rightleftharpoons \isotope{12}{C}$.
 \putfig
     {reactionVectors3}   
     {0pt}
     {\figdn}
     {0.65}
     {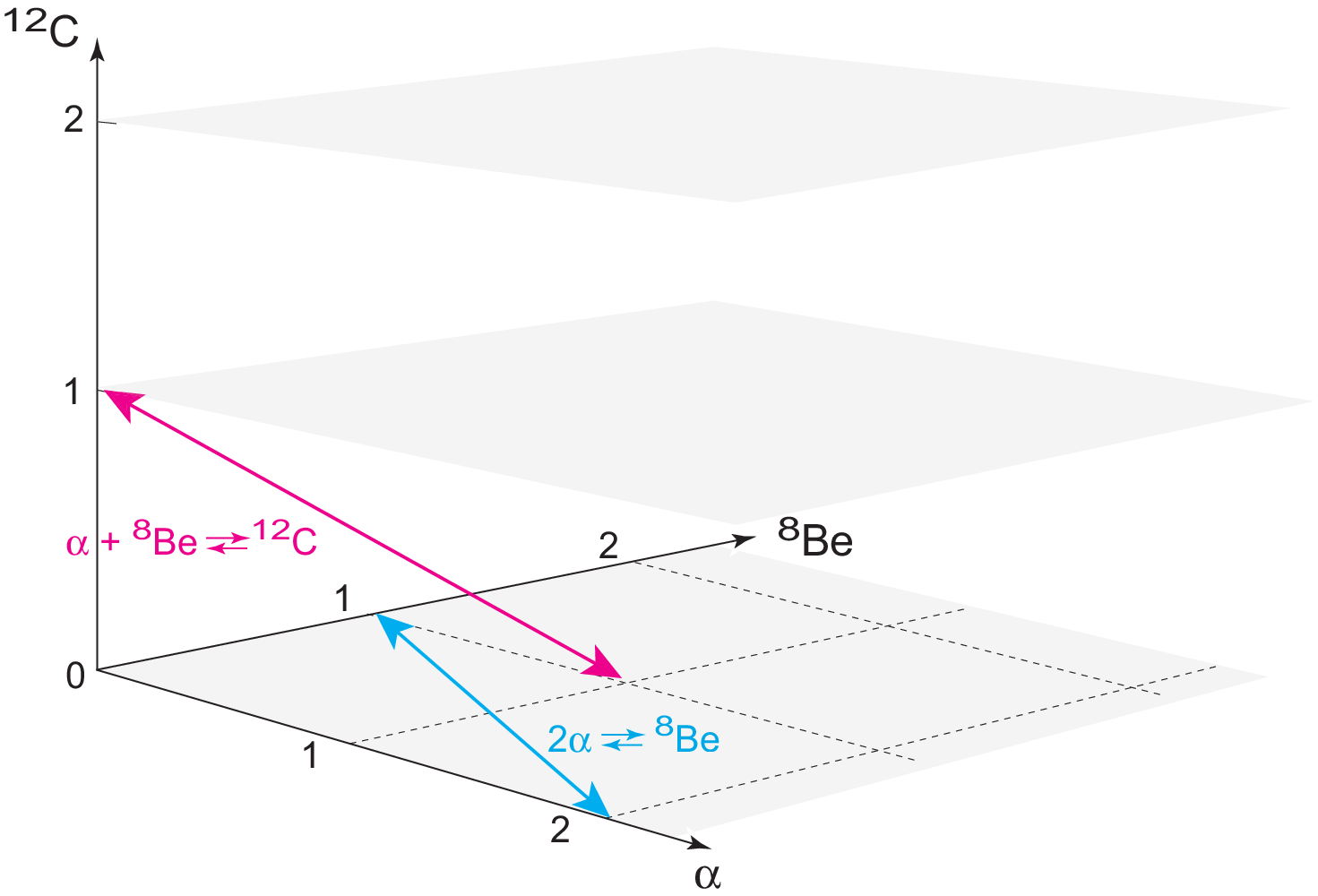}
{(a)~Four reaction vectors for a simple 3-species network corresponding to the
astrophysical triple-$\alpha$ process that converts \isotope{4}{He} to
\isotope{12}{C} in red giant stars.}
Larger networks cannot be visualized so easily, but their algebraic properties
remain completely analogous to those of a simple network like that illustrated
in \fig{reactionVectors3}.

\subsection{Conservation Laws}
\protect\label{conlaws}

Given an initial composition $\bm y_0 = (y^0_1, y^0_2,  \ldots y^0_n)$, a single
pair of reactions labeled by $i$ can produce a composition $\bm y = \bm y_0 +
\alpha_i \bm r_i$, so for a set of $k$ possible reactions
\begin{equation}
 \bm y = \bm y_0 + \sum_{i=1}^k \alpha_i \bm r_i .
\label{vector1.4}
\end{equation}
Define a time-independent vector $\bm c  \in \Phi$ that is orthogonal to each of
the $k$ vectors $\bm r_i$ in \eq{vector1.4}.  Then from \eq{vector1.4}, $\bm c
\cdot \bm y = \bm c \cdot \bm y_0$ and therefore
\begin{equation}
 \sum_{i=1}^n c_i y_i = \sum_{i=1}^n c_i y_0^i = \units{constant},
\label{vector1.5}
\end{equation}
so any vector $\bm c$ orthogonal to the reaction vectors $\bm r_1$, $\bm r_2$,
\ldots $\bm r_k$ gives a linear combination of  abundances that is invariant
under all actions of the network. Equation \eqnoeq{vector1.5} is a {\em
conservation law of the system} that follows purely from the structure of the
network and thus is valid irrespective of dynamical conditions in the network.
The conserved quantities may be determined by solving the equation $\bm r\cdot
\bm c = 0$, where $\bm r$ is a matrix with rows composed of all reaction vectors
\eqnoeq{reactionVector}, for the vectors $\bm c_k$.

Thus, once a basis has been chosen for a specific reaction network it is
straightforward to enumerate all of its possible compositions in terms of a set
of vectors $\bm y_i$, all possible reactions in terms of a set of vectors $\bm
r_j$, and the conservation laws that follow from the structure of the network in
terms of a set of vectors $\bm c_k$. This abstract description of a reaction
network retains a reference to a specific physical problem only through the
choice of basis and the choice of allowed reaction vectors.

\subsection{Reaction Group Classes}
\protect\label{reactionGroupClasses}

It will prove useful to associate inverse reaction pairs in {\em reaction group
classes} ({\em reaction groups,} or {\em RG} for short), in which all reactions
of a group share the same reaction vector $\bm r$, up to a sign. To illustrate
we employ the reaction classifications used in the REACLIB library
\cite{raus2000} that are illustrated in Table \ref{tb:reaclibClasses}.  In this
classification the reactions of importance in nuclear astrophysics are assigned
to eight categories, depending on the number of {\em nuclear species} on the two
sides of the reaction equation.  From this classification  we see that there are
five independent ways to combine the reactions of Table \ref{tb:reaclibClasses}
into reversible reaction pairs, leading to the reaction group classification
illustrated in Table \ref{tb:reactionGroupClasses}.  For example, reaction group
 B consists of reactions from REACLIB reaction class 2 (a $\rightarrow$ b + c)
paired with their inverse reactions (b + c $\rightarrow$ a), which belong to
REACLIB reaction class 4.

\begin{table}[t]
  \vspace{10pt}
  \caption{Reaction classes in the REACLIB \cite{raus2000} library}
  \label{tb:reaclibClasses}\vspace{5pt}
  \begin{small}
    \begin{centering}
      \setlength{\tabcolsep}{5 pt}
      \begin{tabular}{cll}
        \hline
            {Class} &
            {Reaction} &
            {Description or example}

        \\        \hline
            1 &
            a $\rightarrow$ b &
            $\beta$-decay or e$^-$ capture

        \\ 
            2 &
            a $\rightarrow$ b + c &
            Photodisintegration
                                 + $\alpha$ 
	\\ 
            3 &
            a $\rightarrow$ b + c + d&
            $^{12}$C $\rightarrow$ 3$\alpha$ 
	\\ 
            4 &
            a + b $\rightarrow$ c &
            Capture reactions
	\\ 
            5 &
            a + b $\rightarrow$ c + d &
            Exchange reactions
	\\ 
            6 &
            a + b $\rightarrow$ c + d + e &
            $^2$H + $^7$Be $\rightarrow$ $^1$H + 2$^4$He 
	\\ 
            7 &
            a + b $\rightarrow$ c + d + e + f &
            $^3$He + $^7$Be $\rightarrow$ 2$^1$H 
            + 2$^4$He 
	\\ 
            8 &
            a + b + c $\rightarrow$ d \, (+ e) &
            Effective 3-body reactions  
        \\        \hline
      \end{tabular}
\vspace{0pt}
    \end{centering}
  \end{small}
\end{table}

\begin{table}[t]
  \vspace{10pt}
  \caption{Reaction group classes}
  \label{tb:reactionGroupClasses}\vspace{5pt}
  \begin{small}
    \begin{centering}
      \setlength{\tabcolsep}{10 pt}
      \begin{tabular}{cll}
        \hline
            {Class} &
            {Reaction pair} &
            {REACLIB class pairing}

        \\        \hline
            A &
            a $\rightleftharpoons$ b &
            1 with 1

        \\ 
            B &
            a + b $\rightleftharpoons$ c &
            2 with 4
	\\ 
            C &
            a +b + c $\rightleftharpoons$ d&
            3 with part of 8
	\\ 
            D &
            a + b $\rightleftharpoons$ c + d &
            5 with 5
	\\ 
            E &
            a + b $\rightleftharpoons$ c + d  + e&
            6 with part of 8
        \\        \hline
      \end{tabular}
\vspace{10pt}
    \end{centering}
  \end{small}
\end{table}

For each reaction group class the characteristic differential equation governing
the reaction pair considered in isolation is of the form given by \eq{2body1.3},
$dy/dt = ay^2 + by +c$, where $y$ is either an abundance variable (proportional
to a number density) for one of the reaction species, or a progress variable
measuring the change from initial abundances associated with the reaction pair,
and the coefficients $a$, $b$, and $c$ are known parameters depending on the
reaction rates that will be assumed constant within a single network timestep.
An exception occurs for reaction group classes C and E, where there are a 3-body
reactions and the most general form of the differential equation involves cubic
terms,
$
dy/dt = \alpha y^3 + \beta y^2 + \gamma y + \epsilon.
$
These ``3-body'' reactions in astrophysics are typically actually sequential
2-body reactions and we assume that in any timestep $y(t)^3 \simeq y^{(0)}
y(t)^2$, where $y^{(0)}$ is the (constant) value of $y(t)$ at the beginning of
the timestep. This reduces the cubic equation to an effective quadratic equation
of the form (\ref{2body1.3}), with $a = \alpha y^{(0)} + \beta$, $b = \gamma$,
and $c=\epsilon$. This has been found to be a very good approximation in typical
astrophysical environments and we apply it to all 3-body reactions in the
examples discussed here.

\subsection{Equilibrium Constraints}
\protect\label{equilConstraints}

If a reaction pair from a specific reaction group class is near equilibrium,
there will be a corresponding equilibrium constraint that takes the general form
\cite{mott00}
\begin{equation}
 \prod_{i=1}^n y_i^{(b_i-a_i)} = K,
\label{rgclass1.3}
\end{equation}
 where $K$  is a ratio of rate parameters. For example, consider the reaction
group class E  pair $a+b \rightleftharpoons c+d+e$ in isolation, with
differential equations for the populations $y_i$ written in the form
$$
\dot y\tsub a = \dot y\tsub b = -\dot y\tsub c = -\dot y\tsub d
= -\dot y\tsub e = -k\tsub f y_a y_b + k\tsub r y_c y_d y_e.
$$
Then at equilibrium, requiring that the forward flux $-k\tsub f y_ay_b$ and
reverse flux $k\tsub r y\tsub c y\tsub d y\tsub e$ in the reaction pair sum to
zero gives the constraint
$$
\frac{y_a y_b}{y_c y_d y_e} = \frac{k\tsub r}{k\tsub f} \equiv K,
$$
which is of the form \eqnoeq{rgclass1.3}.

\subsection{Systematic Classification of Reaction Group Properties}
\protect\label{RGclassification}

Applying the principles discussed in the preceding paragraphs to the reaction
group classes in Table \ref{tb:reactionGroupClasses} gives the results
summarized for reaction group classes A--E in Appendix
\ref{RGclassificationApp}. This reaction group classification has been developed
assuming astrophysical thermonuclear networks and  a particular parameterization
(REACLIB) of the corresponding reaction rates.  However, the illustrated
methodology is of wider significance. First, since any reaction compilation in
astrophysics could be reparameterized in the REACLIB format, this classification
scheme provides a general partial-equilibrium bookkeeping that could be applied
to any problem in astrophysics. Second, for any large reaction network in any
field we may apply  the classification techniques illustrated here to group all
reactions of the network into reaction group classes and deduce for each
reaction group class analytical expressions for all quantities necessary for
applying a PE approximation. All that is required is to cast the network as a
linear algebra problem by choosing a set of basis vectors corresponding to the
species of the network, and then to define the corresponding reaction vectors of
physical importance within this space. Once that is done, the formalism
developed here may be applied systematically. In principle this classification
need be developed only once for the networks of importance in any particular
discipline, as we have illustrated here specifically for astrophysics.

\subsection{General Methods for Partial Equilibrium Calculations}
\protect\label{generalMethods}

We now have a set of tools to implement partial equilibrium approximations, but
there are some practical issues to resolve before it is possible to make
realistic calculations. To address those issues, let us now outline a specific
application approach. Although our remarks will be illustrated by examples using
astrophysical thermonuclear networks, the methods discussed should be relevant
for a much broader range of problems.

\subsubsection{Overview of Approach}
\protect\label{approachOverview}

We employ the partial equilibrium method in conjunction with the asymptotic
approximation. (There may be advantages in using quasi-steady-state methods
instead of asymptotic ones, but we shall deal with that in future work.) Once
the reactions of the network are classified into reaction groups, the algorithm
has three steps: 
\begin{enumerate}
 \item 
A numerical integration step begins with the full network of differential
equations, but in computing fluxes all terms involving reaction groups judged to
be equilibrated (based on criteria determined by species populations at the end
of the previous timestep) are assumed to sum identically to zero net flux and
are omitted from the flux summations.
\item
A timestep $\Delta t$ is chosen and  used in conjunction with the fluxes to
determine which species satisfy the asymptotic condition. For those species that
are not asymptotic, the change in abundance for the timestep is then computed by
ordinary forward (explicit) finite difference, but for those species that are
judged to be asymptotic the new abundance for the timestep is instead computed
analytically using the asymptotic approximation.
\item
For all species in reaction groups judged to be equilibrated at the beginning of
the timestep, it is assumed that reactions not in equilibrium will have driven
these populations slightly away from equilibrium  during the timestep.  These
populations are then adjusted, subject to the system's conservation laws, to
restore their equilibrium values at the end of the timestep.
\end{enumerate}
Hence the partial equilibrium part of the approximation does not reduce the
number of differential equations integrated numerically within a timestep, but
instead removes systematically the stiffest parts  of their fluxes.  In
contrast, the asymptotic approximation effectively reduces the number of
differential equations integrated numerically by replacing the numerical forward
difference with an analytic asymptotic abundance for those isotopes satisfying
the asymptotic condition.  

The partial equilibrium and asymptotic approaches are complementary because
partial equilibrium can operate microscopically to make the differential
equation for a given isotope less stiff, even if that isotope does not satisfy
the asymptotic condition, while the asymptotic condition removes entire
differential equations from the numerical update and thus can operate
macroscopically to remove stiff reaction components, even if they do not satisfy
partial equilibrium conditions. For brevity we shall often refer simply to the
partial equilibrium (PE) approximation, but  this always means the partial
equilibrium approximation used in conjunction with the asymptotic approximation,
in the manner just described.

\subsubsection{Complications in Realistic Networks}
\protect\label{complexityRealNet}

The complication for the basic approach outlined in \S\ref{approachOverview}
when applied to a realistic network  is that there may be more than one computed
equilibrium value for a given species. This is because the equilibrium abundance
from \eq{2body1.6} is to be computed separately for each reaction group, and a
given isotope often will be found in more than one such group.  In the general
case these computed equilibrium values can be different, since equilibrium
within each reaction group is specified only to the tolerance implied by
$\epsilon_i$ in \eq{2body1.8b}. However, the equilibrium abundances of a given
species computed for each equilibrated reaction group that it is a member of
cannot differ substantially among themselves, since this would contradict the
evidence supplied by the network that the reaction groups are in equilibrium.
For a species $i$ there is only {\em one} actual abundance $y_i$ in the network
at a given time, which must satisfy {\em simultaneously} the equilibrium
conditions for all reaction groups determined to be in equilibrium, within the
tolerances of \eq{2body1.8b}.

Therefore, restoration of equilibrium at the end of a numerical integration
timestep will correspond to setting the abundance of each isotope to a
compromise choice among each of the (similar) predicted equilibrium values for
all equilibrated reaction groups in which it participates. This is a
self-consistent approximation if the variation in possible equilibrium
abundances remains less than the tolerances $\epsilon_i$ used to impose
equilibrium in \eq{2body1.8b}. For the networks that we have tested, this seems
generally to be fulfilled for appropriate choices of $\epsilon_i$.

\subsubsection{Specific Methods for Restoring Equilibrium}

We have investigated three methods for restoring equilibrium at the
end of the numerical timestep:

\begin{enumerate}
 \item 
Reimpose equilibrium ratios (\eq{rgclass1.3}) by Newton-Raphson
iteration.
\item
Reimpose equilibrium abundances (\eq{2body1.6}) by Newton-Raphson
iteration.
\item
Reimpose equilibrium abundances (\eq{2body1.6}) by averaging over progress
variables.
\end{enumerate}
All three methods are described in Ref.\ \cite{guidPE}, where we show that for
the examples investigated to date the third method gives essentially the same
results as the first two and is considerably simpler to implement, since it
involves no matrices and no iteration.  Thus we outline only this
method and refer the reader to Ref.\ \cite{guidPE} for the technical details on
it and the other two methods.

The basic idea is that  in partial equilibrium the isotopic abundances in a
reaction group  evolve according to a single timescale given by \eq{2body1.8},
as discussed in  \S\ref{progvar}. Thus, within a single reaction group,  the
equilibrium abundance of any one isotope, or the progress variable for the
reaction group, determines the equilibrium abundance of all species in the
group.  Furthermore, within a single reaction group the evolution of all species
to equilibrium conserves particle number, by virtue of constraints such as those
of \eq{2body1.1} that are summarized for all five reaction group classes in
Appendix \ref{RGclassificationApp}. Let us exploit this using the progress
variable $\lambda_i$ from each reaction group. If all the reaction groups were
independent, then equilibrium  could be restored by requiring for all reaction
groups in equilibrium $\lambda_i - \bar\lambda_i = 0$, where $\bar\lambda_i$
denotes the equilibrium value of $\lambda_i$ computed from \eq{2body1.6} and
relations like \eq{progress1.5}. Once the equilibrium value of $\lambda_i$ is
computed, the equilibrium values for all other isotopes in the group than follow
from constraints like \eq{progress1.5} that are tabulated for all reaction group
classes in Appendix \ref{RGclassificationApp}.  

The simple considerations of the preceding paragraph are insufficient because
the reaction groups are generally {\em not} independent: the individual network
species may appear in more than one reaction group, as discussed in
\S\ref{complexityRealNet}. The equilibrium condition implies that we must have
approximately equal computed equilibrium values for an isotope participating in
more than one RG, but exact equality does not hold because of the
finite tolerance $\epsilon_i$ used to test for equilibrium in \eq{2body1.8b}.  
Thus, we restore equilibrium for each isotope participating in partial
equilibrium at the end of a timestep by replacing its numerically-computed
abundance with its equilibrium value \eqnoeq{2body1.6}, averaged arithmetically
over all equilibrated reaction groups in which it participates.

There is one final issue:  evolution to equilibrium for individual
reaction groups considered in isolation conserves particle number, but the
averaging procedure introduces a small fluctuation since the average will
generally differ from the individual $\bar y_i$ that were computed conserving
particle number.  Thus, for each timestep, after equilibrium has been restored,
we  rescale all $y_i$  by a multiplicative factor that
restores the total nucleon number to its value at the start of the timestep.

\section{\label{ss:asyTimestep} A Simple Adaptive Timestepper for Explicit
Integration}

For testing the asymptotic and QSS methods a simple adaptive timestepper has
been used that is described in more detail in Ref.\ \cite{guidAsy}:

\begin{enumerate}
 \item 
Compute a trial integration timestep based on limiting the change in populations
that would result from that timestep to a specified tolerance. Choose the
minimum of this trial timestep and the timestep taken in the previous
integration step  and update all populations by the asymptotic or
QSS algorithms. 

\item
Check for conservation of particle number within a specified tolerance range. If
not satisfied, increase or decrease the timestep as appropriate by a small
factor and repeat the timestep using the original fluxes. 
\end{enumerate}
Our asymptotic plus PE calculations use a modified form of this
timestepper that de-emphasizes  the second  step, since the PE algorithm itself
implements approximate probability conservation. This prescription is far from
optimized,  but it gives stable and accurate results for the varied
astrophysical networks that have been tested.  Thus it is adequate for our
primary task here, which is to establish whether explicit methods can even
compete with implicit methods for stiff networks.

\section{Comparisons of Explicit and Implicit Integration Speeds}

In the following examples explicit and implicit integration speeds will be
compared using algorithms at different stages of
development and optimization. Thus, the codes cannot yet be
simply compared head to head.  We shall make a simple  assumption
that for codes at similar levels of optimization the primary difference between
explicit and implicit methods would be in the extra time spent in matrix
operations for the implicit method. Thus, if the fraction of computing time
spent in the linear algebra is $f$ for an implicit code, it will be assumed that
an explicit code at a similar level of optimization could compute a timestep
faster by a factor of $F = 1/(1-f)$. We shall then assume that for implicit and
explicit codes at similar levels of optimization the ratio of speeds for a given
problem would be $F$ times the ratio of integration steps that are required by
the two codes. As discussed later, this may underestimate the speed of a
fully-optimized explicit code relative to implicit codes, but establishes a
lower limit on how fast the explicit calculation can be.

Estimated factors $F$
are shown in Table \ref{tb:explicitSpeedup} for the networks to be discussed,
\begin{table}[t]
  \caption{Speedup factors for explicit vs. implicit timesteps \cite{feg11b}}
  \label{tb:explicitSpeedup}
  \begin{small}
      \setlength{\tabcolsep}{15 pt}
      \begin{tabular}{ccc}
        \hline
            {Network} &
            {Isotopes} &
            {Speedup $F$}

        \\        \hline
            pp &
            7 &
            $\sim 1.5$

        \\ 
            Alpha &
            16 &
            3

        \\
            Nova &
            134 &
            7

        \\
            150-isotope &
            150 &
            7.5

        \\
            365-isotope &
            365 &
            $\sim 20$

        \\        \hline
      \end{tabular}
\vspace{20pt}
  \end{small}
\end{table}
based on data obtained by Feger \cite{feg11b} using the implicit, backward-Euler
code Xnet \cite{raphcode} with both dense solvers (LAPACK \cite{lapack}) and
sparse solvers (MA28 \cite{ma28} and PARDISO \cite{pardiso}), assuming the
optimal solver to be used by the implicit code for a given network. We see that
for very small networks implicit and explicit methods require similar times to
compute a timestep, but for larger networks the explicit computation of a
timestep can be faster than that for the implicit code by a factor of 10 or more
for the networks examined here.

\section{\label{sh:testsAsyQSS} Tests of Asymptotic and QSS Algorithms}

This section presents some representative calculations for astrophysical
thermonuclear networks using the explicit asymptotic (Asy) and
quasi-steady-state (QSS) methods. For astrophysical networks we shall replace
the generic $y_i$ of \eq{equilDecomposition} with population variables common
for astrophysics: the mass fraction $X_i$ or the (molar) abundance $Y_i$, with
\begin{equation}
X_i  = \frac{n_iA_i}{\rho N\tsub A}
\qquad
Y_i \equiv \frac{X_i}{A_i} = \frac{n_i}{\rho N\tsub A}
\label{5.35}
\end{equation}
where $N\tsub A$ is Avogadro's number, $\rho$ is the total mass density, $A_i$
is the atomic mass number, $n_i$ the number density for the species $i$, and
conservation of nucleon number requires $\sum X_i =1$.

\subsection{Explicit Asymptotic and QSS Integration of the pp-Chains}
\protect\label{ppAsy}

The pp-chains that power the Sun provide a spectacular illustration of stiffness
in a simple yet physically-important network. Figure \ref{fig:ppChains}%
 \putfig
     {ppChains}
     {0pt}
     {\figdn}
     {0.97}
     {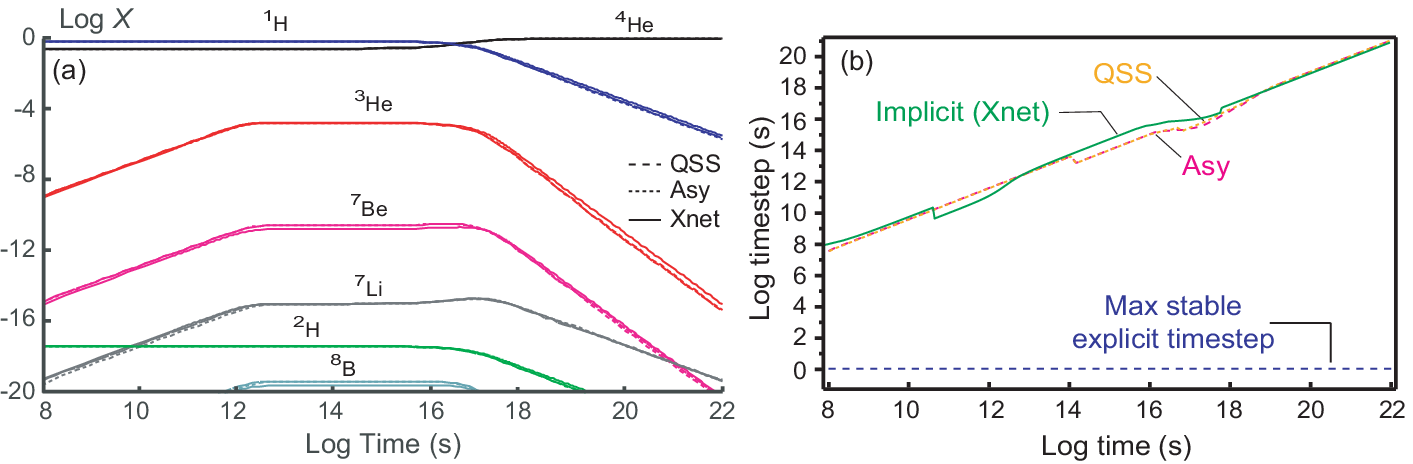}    
{Integration of the pp-chains at constant temperature $T_9 = 0.016$ (where $T_9$
denotes temperature in units of $10^9$ K) and constant density $160
\units{g\,cm}^{-3}$, assuming solar initial abundances. Reaction rates were
taken from the REACLIB library \cite{raus2000}.  (a)~Mass fractions for the
asymptotic method, the QSS method,  and for the standard implicit code Xnet
\cite{raphcode}. (b)~Integration timesteps.}
exhibits integration of the pp-chains at a constant temperature and density
characteristic of the current core of the Sun using the asymptotic method, the
QSS method, and the implicit backward-Euler code Xnet \cite{raphcode}. We see
that the asymptotic and QSS integrations give results for the mass fractions in
rather good agreement with the implicit code over 20 orders of magnitude, and
generally take timesteps $dt \sim 0.1 t$  that are comparable to those for the
implicit code over the entire range of integration. (The asymptotic method
required 333 total integration steps, the QSS method required 286 steps, and the
implicit method required 176 steps for this example.) The maximum stable
timestep for a standard explicit integration method, which typically may be
approximated by the inverse of the fastest rate in the system, is illustrated by
the dashed blue curve in \fig{ppChains}(b). Thus, at the end of the calculation
the explicit integration timesteps are $\sim10^{21}$ times larger than would be
stable in a normal explicit integration. The calculation illustrated in
\fig{ppChains} takes a fraction of a second on a 3 GHz processor with the
asymptotic, QSS, or implicit methods.  In contrast, from \fig{ppChains}(b) we
estimate that a standard explicit method  would require $ \sim 10^{21} \units s$
of processor time to compute the pp-chains to hydrogen depletion, which is a
thousand times longer than the age of the Universe. 

This example is a simple one, but the results  call into question most of what
has been said in the literature concerning the use of explicit methods for stiff
systems.  This network is about as stiff as one will find in any scientific
application, with a maximum integration timestep that is 21 orders of magnitude
larger than the inverse of the fastest rate in the system. Yet both the explicit
asymptotic and explicit quasi-steady-state methods have integrated it with an
efficiency and accuracy comparable to that of a standard implicit algorithm.

\subsection{\label{novaExplosions} Nova Explosions}

The network of the preceding example was extremely stiff, but small. Let us now
examine a case involving an extremely stiff but much larger network. Figure
\ref{fig:nova125D_XplusHydroProfileQSS}(a)%
\putfig
     {nova125D_XplusHydroProfileQSS}
     {0pt}
     {\figdn}
     {1.0}
     {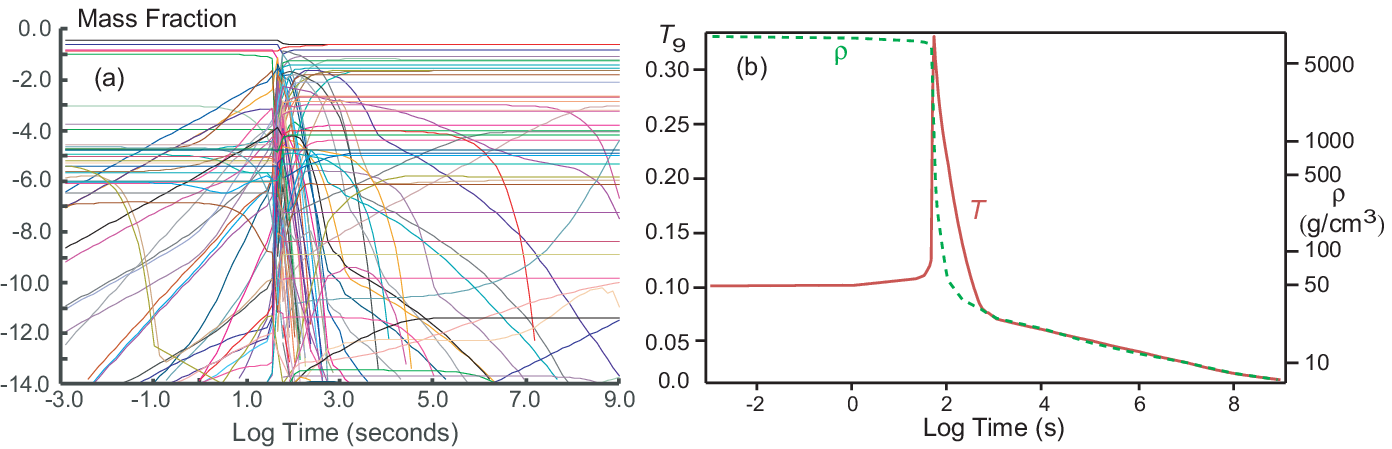}   
{(a)~Mass fractions for a network under nova conditions, corresponding to the
hydrodynamical profile shown in (b). The calculation used the QSS method and a
network containing 134 isotopes coupled by 1531 reactions, with rates taken from
the REACLIB library \cite{raus2000} and initial abundances enriched in heavy
elements \cite{parete-koon03}.}
illustrates a calculation using the explicit QSS algorithm with a hydrodynamical
profile \cite{novaProfile} displayed in \fig{nova125D_XplusHydroProfileQSS}(b)
that is characteristic of a nova outburst. Since there are so many mass-fraction
curves we do not attempt to display a direct comparison with an asymptotic or
implicit calculation, but note that the agreement is rather good, and the total
integrated energy release corresponding to the simulation of
\fig{nova125D_XplusHydroProfileQSS} was within 1\% of that found for the same
network using the  explicit asymptotic approximation. 

The integration timesteps for the calculation in
\fig{nova125D_XplusHydroProfileQSS}(a) are displayed in
\fig{nova125D_dtPlusfractionQSS}(a).%
\putfig
     {nova125D_dtPlusfractionQSS}
     {0pt}
     {\figdn}
     {0.95}
     {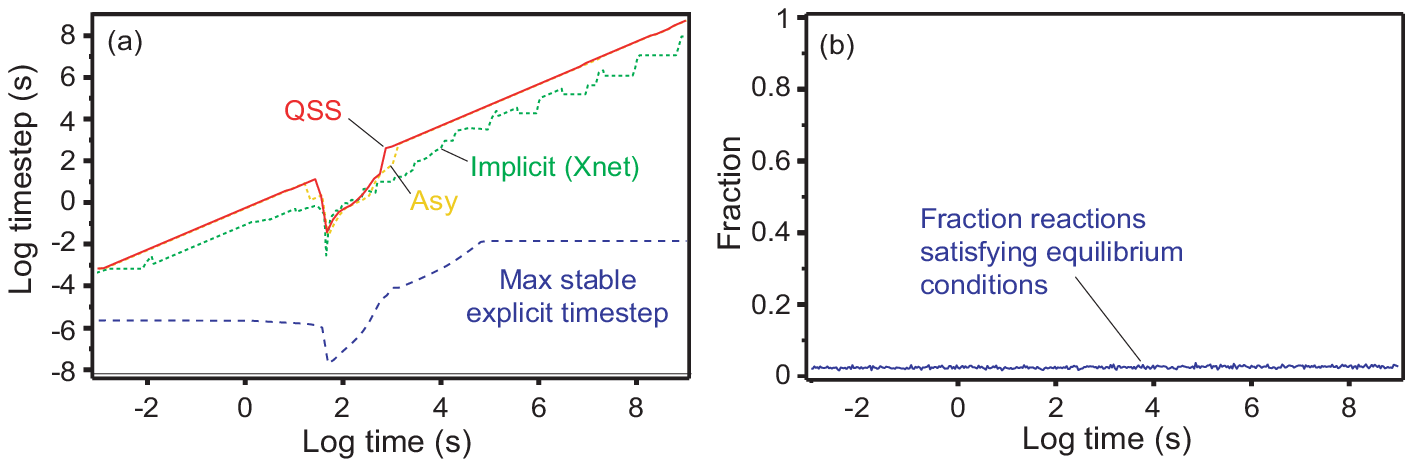}   
{(a)~Timesteps for integration of \fig{nova125D_XplusHydroProfileQSS}.
(b)~Fraction  of reactions that reach partial equilibrium in the QSS
calculation of \fig{nova125D_XplusHydroProfileQSS}.}
Once burning commences, the QSS and asymptotic solvers take timesteps that are
from $10^6$ to $10^{10}$ times larger than would be stable for a normal explicit
integration. The explicit QSS timesteps  illustrated in
\fig{nova125D_dtPlusfractionQSS}(a) are somewhat larger than those of the
asymptotic solver, and comparable to or greater than those for a typical
implicit code: in this calculation the implicit method required 1332 integration
steps, the asymptotic calculation required 935 steps, and the QSS method
required 777 steps. 

Given that for a network with 134 isotopes the explicit codes should be able to
calculate an integration timestep about 7 times faster than the implicit code
because they avoid the manipulation of large matrices (Table
\ref{tb:explicitSpeedup}), these results suggest that the explicit QSS algorithm
is capable of calculating the nova network about 12 times faster and the
explicit asymptotic algorithm about 10 times faster than a state-of-the-art
implicit code. This impressive integration speed for both the QSS and asymptotic
methods applied to a large, extremely stiff network is possible because few
reactions reach equilibrium during the simulation, as determined from
\eq{2body1.8b} and illustrated in \fig{nova125D_dtPlusfractionQSS}(b). 

The ability of explicit methods to deal effectively with large networks under
nova conditions is supported further in work reported by Feger, et al
\cite{feg11b,feg11a}.  There results similar the present ones were demonstrated
with asymptotic methods for a nova simulation using a different network,
different hydrodynamical profile, and different reaction library than those
employed here.

\subsection{\label{tidalSupernova} Tidal Supernova Simulation}

A comparison of asymptotic and QSS mass fractions and timesteps is shown in
\fig{tidalAlphaAsyQSS_X_dt}(a)%
\putfig
{tidalAlphaAsyQSS_X_dt}
{0pt}
{\figdn}
{0.71}
{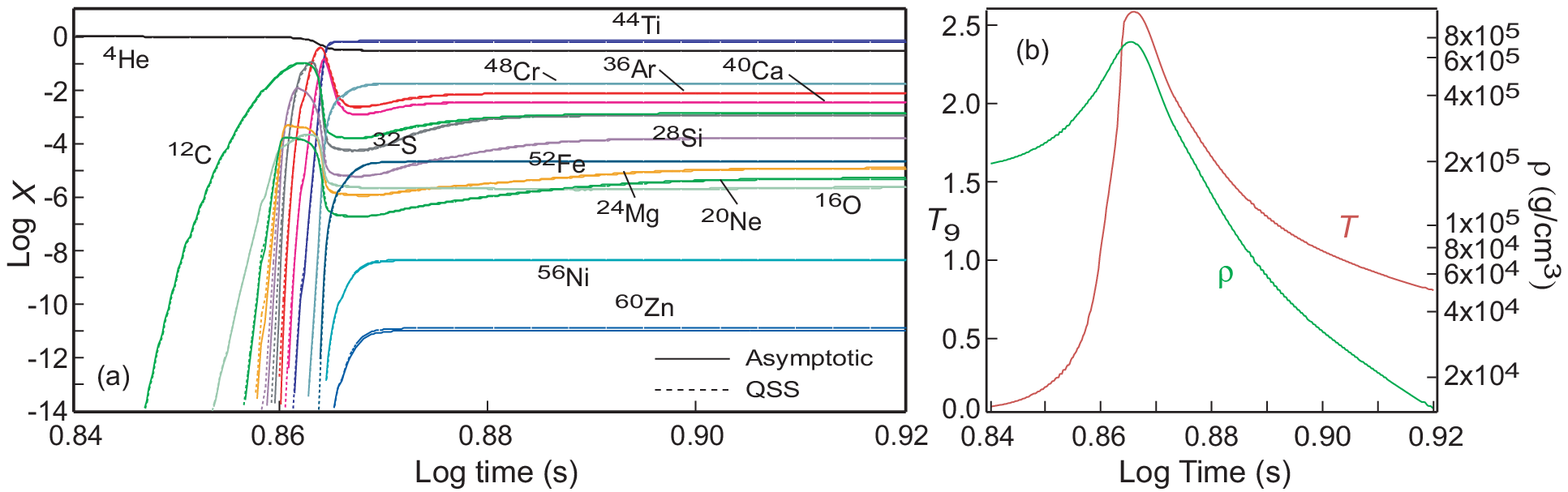} {Comparison of asymptotic and QSS approximation
calculations for an alpha network calculation under tidal supernova conditions.
Initial composition was pure \isotope
4{He} and REACLIB rates \cite{raus2000} were used. (a)~Mass fractions. 
(b)~Hydrodynamical profile \cite{tidalSupernova}.  }
for an alpha network with a hydrodynamical profile illustrated in
\fig{tidalAlphaAsyQSS_X_dt}(b) that is characteristic of a supernova induced by
tidal interactions in a white dwarf \cite{tidalSupernova}. The integration
timestepping  is compared for QSS, an asymptotic calculation, and the implicit
code Xnet in \fig{tidalAlpha_dtPlusFractionQSS}(a).
\putfig
     {tidalAlpha_dtPlusFractionQSS}
     {5pt}
     {\figdn}
     {0.90}
     {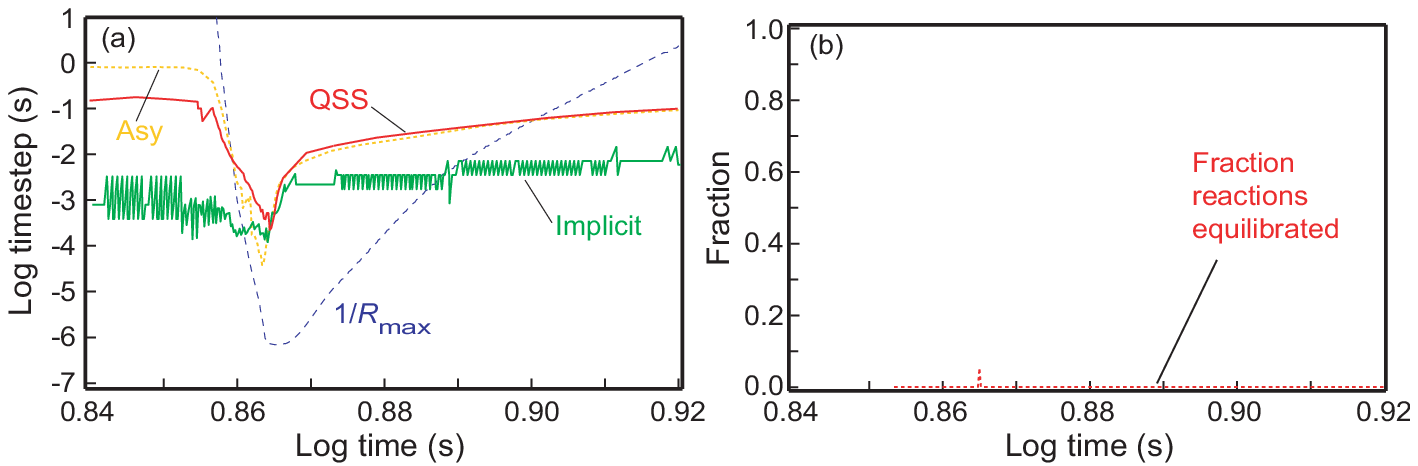} 
{(a)~Integration timesteps and maximum stable purely-explicit timestep
$\sim 1/R\tsub{max}$ for the calculation in \fig{tidalAlphaAsyQSS_X_dt}.
(b)~Fraction of isotopes that become asymptotic and fraction of reactions
equilibrated in the network for the calculation in \fig{tidalAlphaAsyQSS_X_dt}.}
We see that the timestepping for the QSS calculation is somewhat better than for
the asymptotic code and considerably better than for the implicit code (242
total integration steps for the QSS calculation, 480 steps for the asymptotic
calculation, and  1185 steps for the implicit calculation).  Since from Table
\ref{tb:explicitSpeedup}  an optimized explicit code could compute timesteps
about 3 times as fast as an implicit code for an alpha network, we may estimate
that the QSS code is capable of calculating this network perhaps 15 times
faster, and the asymptotic code  perhaps 7 times faster, than the implicit code.
The relatively good timestepping for the QSS and asymptotic methods in this case
again is primarily because almost no reactions in the network come into
equilibrium over the full range of the calculation, as illustrated in
\fig{tidalAlpha_dtPlusFractionQSS}(b).

A calculation for the hydrodynamical profile illustrated in
\fig{tidalAlphaAsyQSS_X_dt}(b) for a 365-isotope network using asymptotic and
implicit methods is illustrated in \fig{tidal365Xdt}.%
\putfig
     {tidal365Xdt}
     {0pt}
     {\figdn}
     {0.72}
     {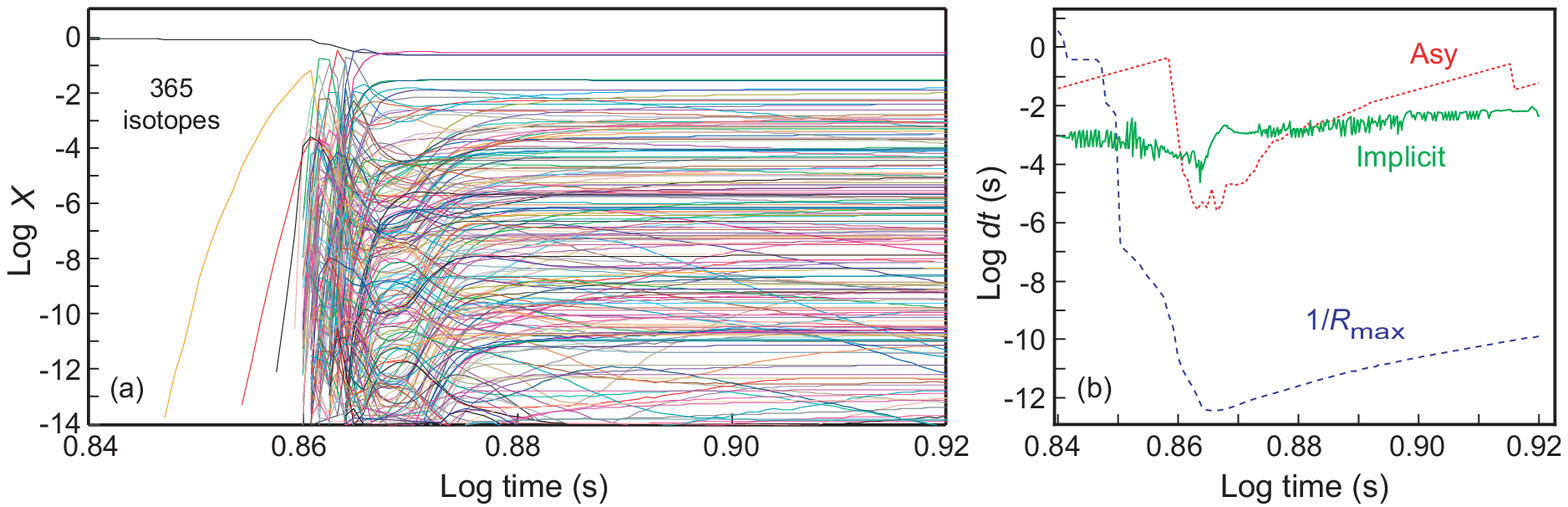}
{(a)~Mass fractions calculated in asymptotic approximation using asymptotic
methods for a 365-element network with 4325 reaction couplings under tidal
supernova conditions, corresponding to the hydrodynamical profile shown in
\fig{tidalAlphaAsyQSS_X_dt}(b).  Rates were taken from REACLIB \cite{raus2000}
and
the initial composition was pure \isotope{4}{He}. (b)~The corresponding
integration timesteps and maximum stable explicit timestep $\sim
1/R\tsub{max}$.}
The implicit calculation
required 1455 integration steps, compared
with 5778 steps for the asymptotic calculation.  But Table
\ref{tb:explicitSpeedup} indicates that for this 365-isotope network the
explicit code can calculate each timestep about 20 times faster than the
implicit code, so an optimized asymptotic code should be capable of performing
the integration in \fig{tidal365Xdt} perhaps 5 times faster than a
state-of-the-art implicit code.

This ability of explicit methods to deal effectively with large networks under
tidal supernova conditions is supported further by results from
Refs.~\cite{feg11b,feg11a}. Although different networks and different reaction
network rates were used in these references, the explicit asymptotic method was
again found to be highly competitive with standard implicit methods for the
tidal supernova problem.

\subsection{\label{sh:noncompetive} Non-Competitive Explicit Timesteps in the
Approach to Equilibrium}

The examples shown to this point have deliberately emphasized examples
from networks in which few reactions have become microscopically equilibrated.
For such cases the estimated integration speed for optimized quasi-steady-state
and asymptotic explicit methods is often comparable to, and in some cases may
exceed, that for current implicit codes.  Let us now turn to an example
representative of a whole class of networks where this is decidedly not true.
The calculation in \fig{compare_dt_asy_qssT9_5rho1e8Alpha}%
\putfig
{compare_dt_asy_qssT9_5rho1e8Alpha}
{0pt}
{\figdn}
{0.70}
{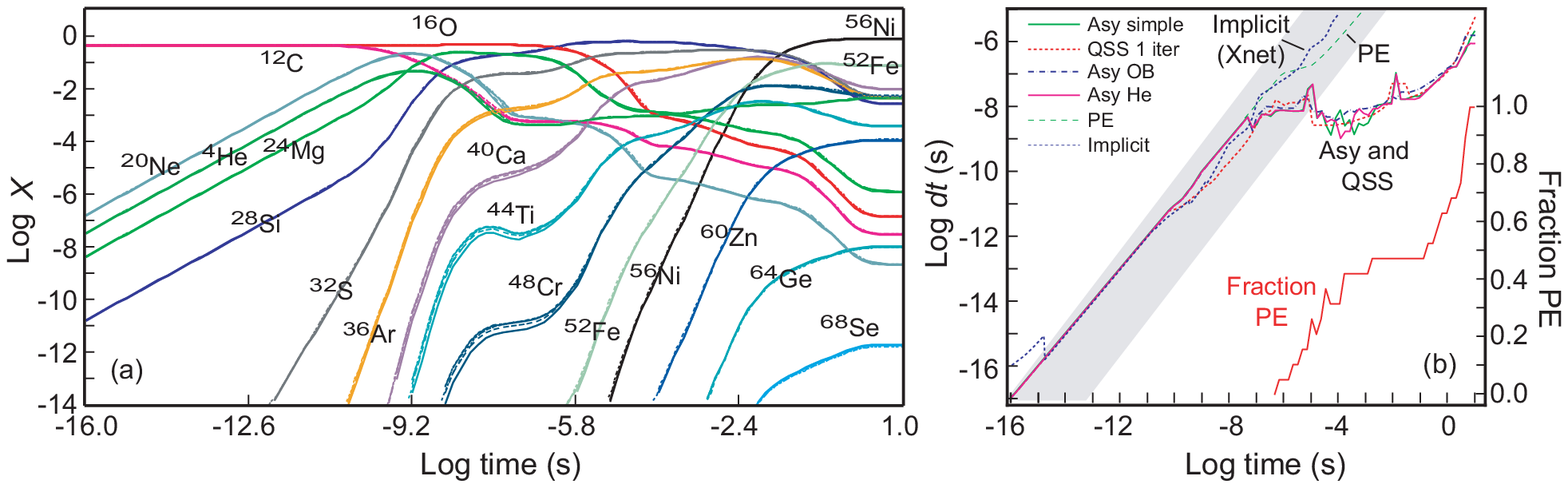}
{Comparison of asymptotic and QSS approximations for an alpha network with
constant temperature $T_9 = 5$ and constant density of $10^8
\units{g\,cm}^{-3}$, using REACLIB rates \cite{raus2000} and initial equal mass
fractions of \isotope{12}{C} and \isotope{16}{O}.   (a)~Isotopic mass fractions.
(b)~Integration timesteps (left axis) and fraction of reaction in partial
equilibrium (right axis). The gray shaded area represents roughly the region
that the explicit timestep profile must lie in to have a chance to compete with
implicit methods. The different asymptotic methods are labeled Asy and are
described in Ref.\ \cite{guidAsy}. The dashed green line (PE) represents the
timestepping using the partial equilibrium methods that will be discussed in
\S\ref{sh:testsAlpha}. }
compares various asymptotic approximations and a QSS calculation with an
implicit calculation for an alpha-particle network (see Table
\ref{tb:alphaNetworkFullReverse}) at a constant temperature and density that
might be found in a deflagration-burning zone for a Type Ia supernova
simulation. Two important conclusions follow from these results.

\begin{enumerate}
 \item 
All of the QSS and asymptotic cases shown are similar, with integrated final
energies (not shown) that lie within 1\% of each other and total integration
times that lie within 25\% of each other.
\item
The asymptotic and QSS methods all give timesteps that are potentially
competitive with implicit methods (they lie in the shaded gray area) at earlier
times, but fall far behind for times greater than about $10^{-5}$ s. 
\end{enumerate}
The reason for conclusion (2) may be seen from the solid red curve on the right
of \fig{compare_dt_asy_qssT9_5rho1e8Alpha}(b), which represents the fraction of
reactions  that satisfy partial equilibrium conditions. Asymptotic and
quasi-steady-state approximations work very well as long as the network is
well-removed from equilibrium, but as soon as significant numbers of reactions
become equilibrated the asymptotic and QSS timestepping begins to lag by large
margins. As we now show, {\em partial equilibrium methods} must be used in
conjunction with asymptotic or QSS methods to recover competitive timestepping
in the approach to equilibrium.  A preview of those results is displayed in
\fig{compare_dt_asy_qssT9_5rho1e8Alpha}(b). The dashed green curve labeled PE
corresponds to an asymptotic plus partial equilibrium approximation that
exhibits highly-competitive timestepping relative to that of the implicit
calculation, even as the network approaches equilibrium.

\section{\label{sh:testsAlpha} Tests of Partial Equilibrium for Some
Thermonuclear Alpha Networks}

The  partial equilibrium algorithm of \S\ref{partialEq} has been tested in a
variety of thermonuclear alpha networks. This section gives some representative
examples of those calculations. Because they are among the toughest numerical
problems for reaction networks, we shall concentrate on conditions expected in
Type Ia supernova explosions (temperatures from $10^7$--$10^{10}$ K and
densities from $10^7$--$10^9$ g\,cm$^{-3}$). Since the Type Ia explosion is
triggered by a thermonuclear runaway in the degenerate matter of a white dwarf,
the reaction network must deal with temperatures that can change at rates in
excess of $10^{17}$ K/s.  Such conditions lead to rapid equilibration in systems
of extremely high stiffness and represent a demanding test of any numerical
integration method. Although our longer-term goal is application of the present
methods to larger networks with hundreds or even thousands of isotopes, an alpha
network provides a fairly realistic, highly-stiff system for initial tests that
has the advantage of being small enough to provide transparency in how the
algorithm functions. It should also be noted that the most ambitious
calculations to date for astrophysical thermonuclear networks coupled to
hydrodynamical simulations have employed alpha networks (or even more schematic
ones). Reactions and corresponding reaction groups  are displayed in Table
\ref{tb:alphaNetworkFullReverse}.
{\renewcommand\arraystretch{1.10}
\begin{table}[t]
  \centering
  \caption{Reactions of the alpha network for partial equilibrium calculations.
The reverse reactions such as $\isotope{20}{Ne} \rightarrow
\alpha+\isotope{16}{O}$  are photodisintegration reactions, $\gamma +
\isotope{20}{Ne} \rightarrow \alpha+\isotope{16}{O}$, with the photon $\gamma$
suppressed in the notation since we track only nuclear species in the network.}
  \label{tb:alphaNetworkFullReverse}
\vspace{8pt}
  \begin{small}
    \begin{centering}
      \setlength{\tabcolsep}{5 pt}
      \begin{tabular}{cccc}
        \hline
	Group & Class & Reactions & Members
	\\ \hline
            1 &
            C &
            $3\alpha\rightleftharpoons\isotope{12}{C}$ &
            4
	\\
	2 &
            B &
            $\alpha+\isotope{12}{C}\rightleftharpoons\isotope{16}{O}$ &
            4
\\
	3 &
            D &
           
$\isotope{12}{C}+\isotope{12}{C}\rightleftharpoons\alpha+\isotope{20}{Ne}$ &
            2
\\
	4 &
            B &
            $\alpha+\isotope{16}{O}\rightleftharpoons\isotope{20}{Ne}$ &
            4
\\
	5 &
            D &
           
$\isotope{12}{C}+\isotope{16}{O}\rightleftharpoons\alpha+\isotope{24}{Mg}$ &
            2
\\
	6 &
            D &
           
$\isotope{16}{O}+\isotope{16}{O}\rightleftharpoons\alpha+\isotope{28}{Si}$ &
            2
\\
	7 &
            B &
            $\alpha+\isotope{20}{Ne}\rightleftharpoons\isotope{24}{Mg}$ &
            4
\\
	8 &
            D &
           
$\isotope{12}{C}+\isotope{20}{Ne}\rightleftharpoons\alpha+\isotope{28}{Si}$ &
            2
\\
	9 &
            B &
            $\alpha+\isotope{24}{Mg}\rightleftharpoons\isotope{28}{Si}$ &
            4
\\
	10 &
            B &
            $\alpha+\isotope{28}{Si}\rightleftharpoons\isotope{32}{S}$ &
            2
\\
	11 &
            B &
            $\alpha+\isotope{32}{S}\rightleftharpoons\isotope{36}{Ar}$ &
            2
\\
	12 &
            B &
            $\alpha+\isotope{36}{Ar}\rightleftharpoons\isotope{40}{Ca}$ &
            2
\\
	13 &
            B &
            $\alpha+\isotope{40}{Ca}\rightleftharpoons\isotope{44}{Ti}$ &
            2
\\
	14 &
            B &
            $\alpha+\isotope{44}{Ti}\rightleftharpoons\isotope{48}{Cr}$ &
            2
\\
	15 &
            B &
            $\alpha+\isotope{48}{Cr}\rightleftharpoons\isotope{52}{Fe}$ &
            2
\\
	16 &
            B &
            $\alpha+\isotope{52}{Fe}\rightleftharpoons\isotope{56}{Ni}$ &
            2
\\
	17 &
            B &
            $\alpha+\isotope{56}{Ni}\rightleftharpoons\isotope{60}{Zn}$ &
            2
\\
	18 &
            B &
            $\alpha+\isotope{60}{Zn}\rightleftharpoons\isotope{64}{Ge}$ &
            2
\\
	19 &
            B &
            $\alpha+\isotope{64}{Ge}\rightleftharpoons\isotope{68}{Se}$ &
            2
        \\        \hline
      \end{tabular}
\vspace{20pt}
    \end{centering}
  \end{small}
\end{table}
}
REACLIB \cite{raus2000} was used for all rates except that inverse rates for
reaction groups 3, 5, and 6 are not included in the standard REACLIB tabulation
and were taken from Ref.\ \cite{JINA}. For all calculations the equilibrium
criteria were imposed using \eq{2body1.8b}, with a constant value $\epsilon_i =
0.01$ for the tolerance parameter.

\subsection{Example at Constant Temperature and Density}

A calculation for constant $T_9=5$ and density of $1\times 10^{7}
\units{g\,cm}^{-3}$ in an alpha network is shown in \fig{alpha507PE}.
 \putfig
     {alpha507PE}
     {0pt}
     {20pt}
     {0.75}
     {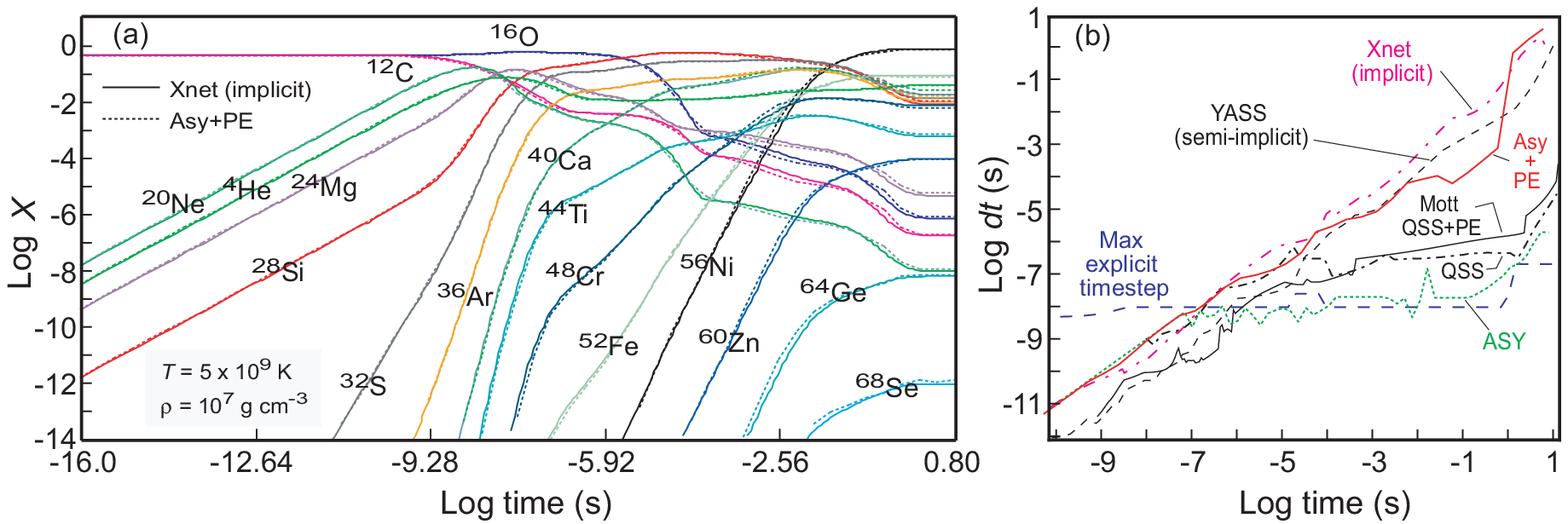}
{Calculation for constant $T= 5 \times 10^9 \units K$ ($T_9=5$) and
$\rho=1\times 10^{7}\units{g\,cm}^{-3}$ alpha network (16 isotopes, 48
reactions, and 19 reaction groups). Initial equal mass fractions of
\isotope{12}{C} and \isotope{16}{O}, and rates from REACLIB \cite{raus2000} and
Ref.\ \cite{JINA} were used. (a)~Mass fractions. (b)~Integration timesteps for
implicit code Xnet \cite{raphcode}, semi-implicit code YASS \cite{YASS}
(reproduced from from Ref.\ \cite{mott99}), asymptotic plus PE (present work),
QSS plus PE calculation reproduced from Ref.\ \cite{mott99}, QSS (present work),
and asymptotic (present work). }
The calculated asymptotic plus partial equilibrium (Asy + PE) mass fractions are
compared with those of an implicit code in \fig{alpha507PE}(a). There are small
discrepancies in localized regions, especially for some of the weaker
populations, but overall agreement is rather good. Mass fractions down to
$10^{-14}$ are displayed for reference purposes.  However, for reaction networks
coupled to hydrodynamics only mass fractions larger than say $\sim
10^{2}-10^{-3}$ are likely to have significant influence on the hydrodynamics. 
Thus, the largest discrepancies between PE and implicit mass fractions in
\fig{alpha507PE}(a), and in the other examples that will be discussed, imply
uncertainties in the total mass being evolved by the network that would be
irrelevant in a coupled hydrodynamical simulation.

Timestepping for various integration methods is illustrated in
\fig{alpha507PE}(b). The PE calculation required 3941 total integration steps
while the implicit code required only 600 steps, but this factor of 6.5
timestepping advantage is offset substantially by the expectation that for a
16-isotope network an explicit calculation should be about 3 times faster than
the implicit calculation for each timestep (Table \ref{tb:explicitSpeedup}).
Thus we conclude that for fully-optimized codes the implicit calculation would
be perhaps twice as fast for this example. Since the semi-implicit timestepping
curve was reproduced from another reference the exact number of integration
steps is not available, but a comparison of the curves in \fig{alpha507PE}(b)
suggests that an optimized partial equilibrium code is likely to be at least as
fast as the semi-implicit YASS code for this example.

The Asy + PE and implicit timesteps are  many orders of magnitude
better at late times than those of the purely asymptotic (labeled Asy) and
purely quasi-steady-state (labeled QSS) calculations. The reason is clear
from \fig{mottCompare507fractionShort}.
 \putfig
     {mottCompare507fractionShort}
     {0pt}
     {\figdn}
     {0.45}
     {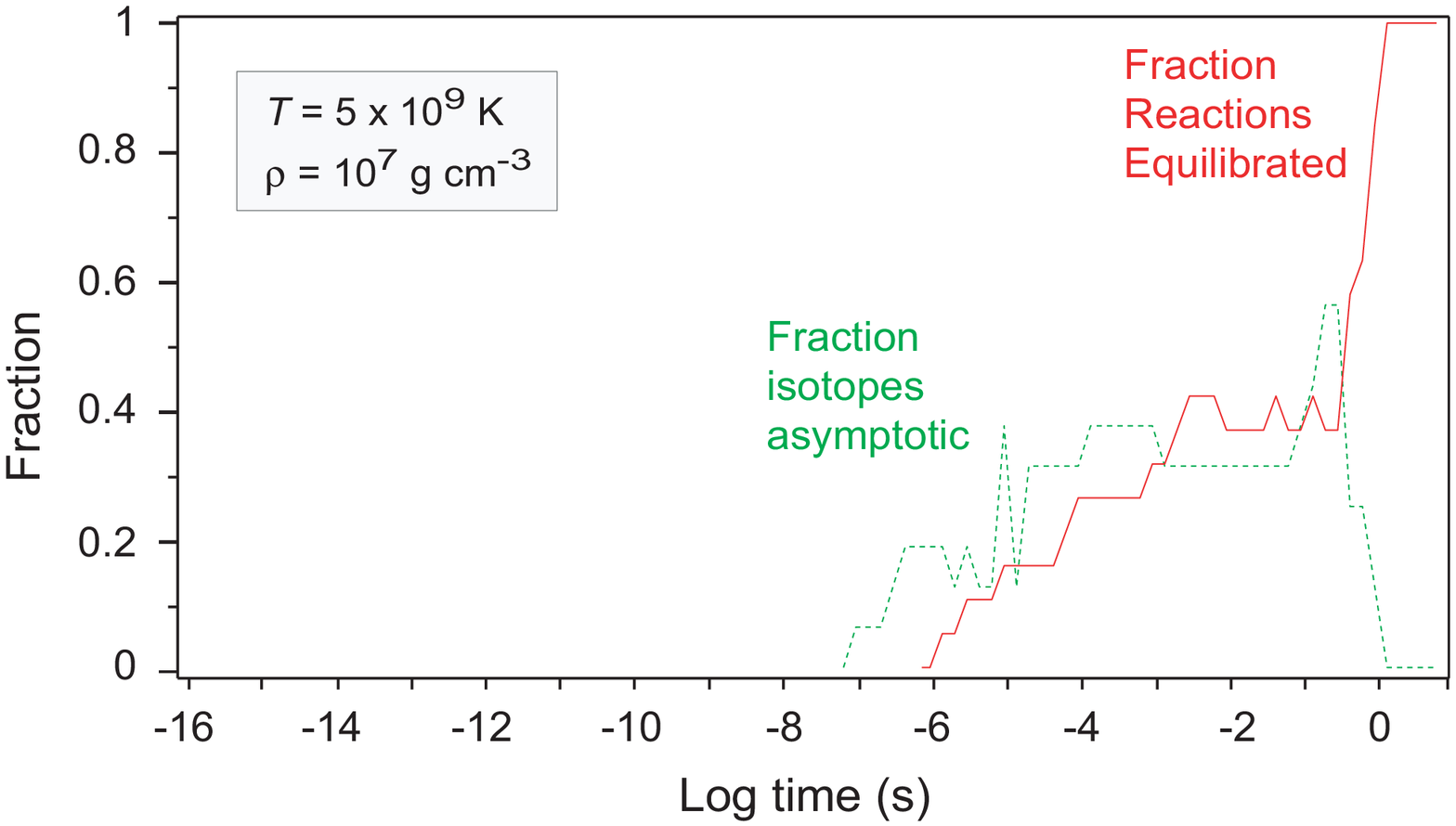}
{Fraction of reactions that are treated as being in equilibrium  as a function
of integration time for the calculation of \fig{alpha507PE}. 
Also shown is the fraction of isotopes that become asymptotic in the PE
calculation.}
For times later than $\sim\log t = -6$, significant numbers of reaction groups
come into equilibrium [as determined by the criteria of \eq{2body1.8b}], which
asymptotic and QSS methods alone are not designed to handle. The earlier
application by Mott \cite{mott99} of a QSS plus PE calculation for this same
system is seen to lag  behind the current implementation of asymptotic plus
partial equilibrium in timestepping at late times by a factor of 1000 or more.
In fact, the timestepping reproduced from Ref.\ \cite{mott99} for QSS + PE is
only a little better than that of the pure QSS results from the present paper,
and clearly is not competitive with that of either the implicit or semi-implicit
calculations, or the present asymptotic plus partial equilibrium result.
Thus we see in this example that the huge speed advantage of implicit methods
relative to pure asymptotic or quasi-steady-state methods in the approach to
equilibrium that was demonstrated in \S\ref{sh:noncompetive} has been
essentially erased by a proper treatment of partial equilibrium in the explicit
integrations. In addition, the present implementation of asymptotic plus PE
methods is seen to be much faster than previous applications of
explicit partial equilibrium methods to this network.

\subsection{Example with a Hydrodynamical Profile}

The preceding example employed an alpha network at extreme but constant
temperature and density. A partial equilibrium calculation using a
hydrodynamical profile with the dramatic temperature rise characteristic of a
burning wave in a Type Ia supernova simulation is illustrated in
\fig{viktorExtended2AlphaCompositePE}.%
\putfig
{viktorExtended2AlphaCompositePE}
{0pt}
{\figdn}
{0.95}
{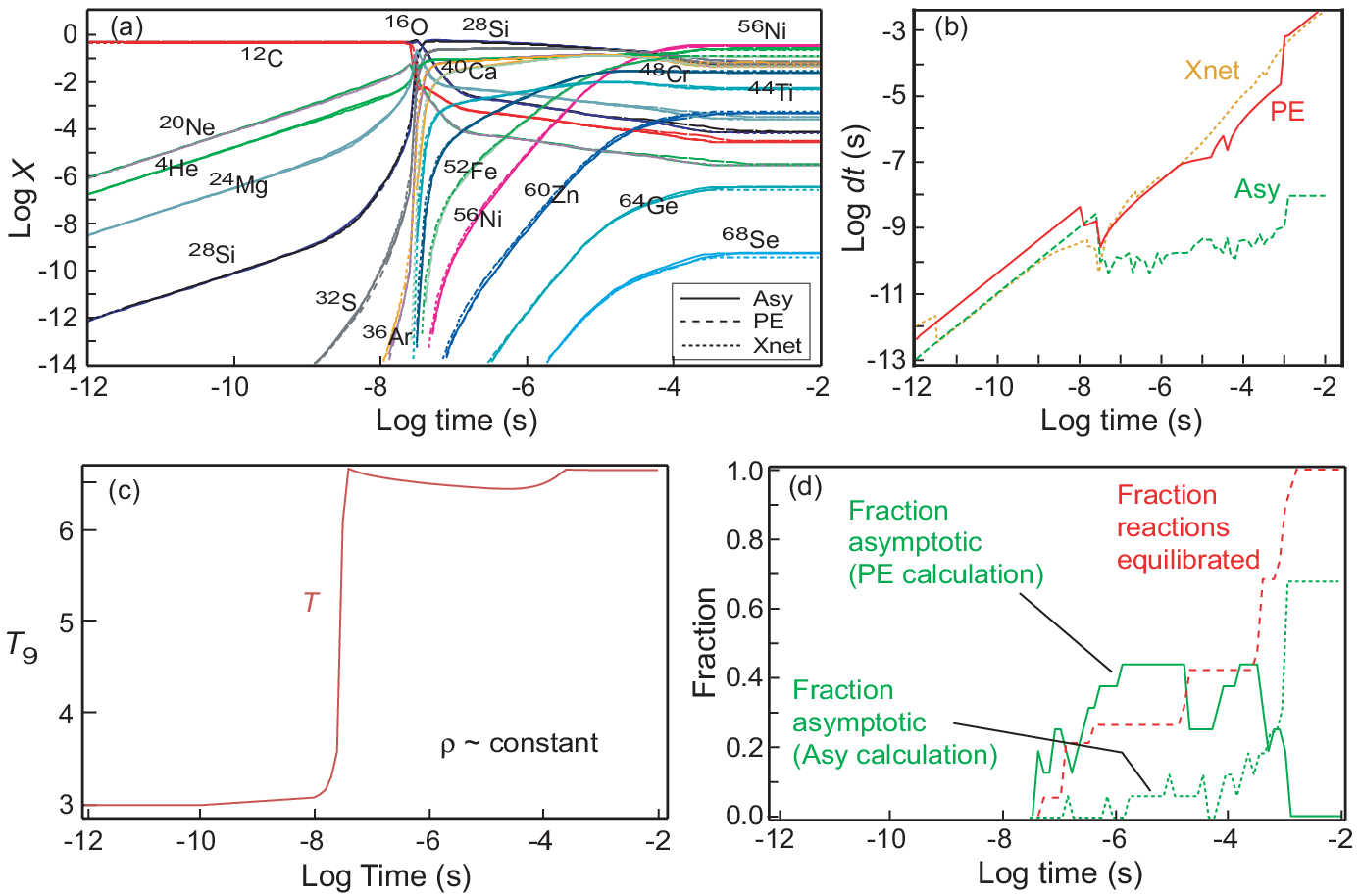}
{Asymptotic, partial equilibrium, and implicit calculations for an alpha network
under Type Ia supernova conditions.  (a)~Mass fractions. (b)~Integration
timesteps. (c)~The hydrodynamic profile. (d)~Fraction of isotopes that become
asymptotic and fraction of network reactions that become equilibrated. Solid
green indicates the fraction of asymptotic isotopes in the PE calculation;
dashed green indicates the fraction of asymptotic isotopes in the purely
asymptotic calculation.  }
These are challenging conditions for the reaction network. During the
thermonuclear runaway the temperature   increases by 3.6 billion K in only $2
\times 10^{-8}$ s, corresponding to a rate of temperature change $1.7 \times
10^{17} \units {K/s}$, and the fastest and slowest rates in the network differ
by approximately 10 orders of magnitude. Again we see that the partial
equilibrium timestepping is competitive with that of the implicit code.  The
partial equilibrium method required 596 total integration timesteps and the
implicit code required 553 steps, so an optimized partial equilibrium
calculation would be perhaps three times faster than the corresponding implicit
calculation because of its speed advantage in computing each timestep. At late
times the timestepping for both the implicit and partial equilibrium
calculations is again found to be orders of magnitude larger than that from the
purely asymptotic calculation. The reason can be seen from
\fig{viktorExtended2AlphaCompositePE}(d): at late times the network is very
strongly equilibrated and can be integrated efficiently with an explicit method
only if the equilibrating reactions are identified and removed from the
numerical integration. 

The synergism of the asymptotic and partial equilibrium approximations is
illustrated by \fig{viktorExtended2AlphaCompositePE}(d), where we see that in
the asymptotic plus PE calculation isotopes become asymptotic much earlier than
in the purely asymptotic calculation. Thus the much larger timestep that the PE
plus asymptotic calculation takes relative to the purely asymptotic calculation
in the time interval $\log t \sim -7$ to $\log t \sim -3$ is enabled by a
combination of reduced stiffness because of the PE approximation and reduced
stiffness because more isotopes become asymptotic at early times.  This
synergism is common and we find typically that replacing the stiffest parts of a
network by a complementary set of algebraic constraints exploiting both
asymptotic and partial equilibrium conditions is much more effective than either
set of constraints used alone.

\subsection{Synopsis of Partial Equilibrium Results}

The examples shown above are representative of various problems that have been
investigated with the new asymptotic plus PE methods. Our general conclusion is
that for alpha networks under the extreme conditions corresponding to a Type Ia
supernova explosion, the asymptotic plus partial equilibrium algorithm is
capable of timestepping that is orders of magnitude better than asymptotic or
QSS approximations alone when the system approaches equilibrium. These timesteps
typically lie in the same ballpark as those from current implicit and
semi-implicit codes, once the partial equilibrium algorithm has removed the fast
timescales associated with approach to equilibrium.  Because the explicit
methods can compute the timestep more efficiently, we find that in many cases
they project to be at least as fast as current implicit codes, even for the
relatively small networks used as examples here.

\section{\label{improveSpeed} Improving the Speed of Explicit Codes}

The preceding discussion has assumed that the relative speed of explicit and
implicit methods can be compared---even if they are presently implemented with
different levels of optimization---by comparing the ratio of integration
timesteps required to do a particular problem with each method multiplied by a
ratio of times to execute a step with each method. The assumption of this
comparison is that once the matrix overhead associated with implicit methods is
factored out, the rest of the work required in a timestep is similar for
explicit and implicit methods. This is useful for a first estimation, and is
sufficient for the primary purpose of this paper, which is to determine whether
explicit methods can take stable timesteps that are even in the same ballpark as
implicit methods. However, the results presented here should not be
overinterpreted.  Because of different levels of optimization for present
implicit and explicit timestepping algorithms, and possible differences
resulting from choices of integration tolerance parameters for implicit and
explicit codes, the present comparisons of integration timesteps are probably
uncertain by at least an order of magnitude. Thus, the essential message of the
present results is that, contrary to most previous claims, explicit methods can
compete favorably with implicit methods for a range of extremely stiff problems.
A definitive comparison of whether implicit or explicit methods are faster for
specific problems, and by how much, must await more complete development and
optimization for the new explicit methods.

As has been noted, the explicit timestepping algorithm employed in this paper is
serviceable but is not very optimized. Thus, it is possible that the number of
integration steps required by our explicit methods has been overestimated for
the examples that have been discussed. But in addition to this obvious potential
for optimization, there are at least two additional reasons why we may not yet
be realizing the full capability of explicit methods.

(1)~For systems near equilibrium, this paper has emphasized partial equilibrium
methods in conjunction with asymptotic approximations.  But we have also
presented evidence that quasi-steady-state (QSS) methods give results comparable
to asymptotic methods, often with timesteps that can be somewhat larger. Thus,
it is possible that a QSS plus partial equilibrium approximation could give even
better timestepping than the asymptotic plus partial equilibrium examples shown
here. This possibility has not yet been investigated. 

(2)~Potentially of most importance, the observation that
algebraically-stabilized explicit methods may be viable for large,
extremely-stiff networks changes the rules of optimization for large reaction
networks. For standard implicit methods, the most effective optimizations
improve the numerical linear algebra, since in large networks the bulk of the
computing time for implicit methods is spent in matrix inversions. But with
algebraically-stabilized explicit methods there are no matrix inversions and the
majority of the time is spent in computing the rates at each step. Faster
computation of reaction rates has only a small impact on the speed of an
implicit code for large networks, so there has been little incentive to worry
about this before. But now we see that our new explicit methods can gain even
more in speed relative to implicit methods by increasing (through software or
hardware) the efficiency for computing rates at each timestep. These potential
speed gains are separate from, and in addition to, those accruing from the
absence of matrix inversions in explicit methods that have been emphasized in
preceding sections.

Let us cite a simple example illustrating this second point.  We have found in a
single-zone hydrodynamics calculation under Type Ia supernova conditions that
using a simple rate-interpolation scheme to avoid recomputing rates for every
hydrodynamical timestep permits the speed for explicit asymptotic integration
for a 150-isotope network to be increased by a factor  $\sim 20$ in the region
of strong silicon burning. 

Also of interest in this context is whether the architecture of modern multicore
processors with associated GPU accelerators can be exploited to compute rates
more efficiently for large-scale simulations of reaction networks coupled to
fluid dynamics. We reiterate that such optimizations could be used for implicit
integrations too, but they will not increase their speeds by as much because
rates are a smaller part of the computing budget for implicit integration of a
large network.

\section{\label{sh:extendFull} Extension of Partial Equilibrium to Larger
Thermonuclear Networks}

Well away from equilibrium, asymptotic and QSS approximations can
compete with current implicit codes even for extremely stiff systems. There are
important cases where the system does not equilibrate strongly during the
evolution of most physical interest (for example, those described in
\S\S\ref{ppAsy}--\ref{tidalSupernova}). Even in problems where equilibrium is
important overall, there will be many hydrodynamical zones and timesteps for
which equilibrium plays a small role. In these situations, the asymptotic and
QSS approximations alone may be adequate to solve large networks efficiently.
However, to compete systematically with implicit solvers across a broad range of
problems, asymptotic and QSS approximations must be augmented by PE methods to
remain viable in those zones where equilibrium is important.

The examples shown here have demonstrated that asymptotic (and presumably QSS)
plus partial equilibrium methods can solve stiff thermonuclear alpha networks
with timestepping similar to that for implicit or semi-implicit codes, and
accuracy more than sufficient to couple such an algorithm to fluid dynamics
simulations. Explicit methods scale linearly and hence more favorable with
network size than implicit methods, so their relative advantage grows for larger
networks. Thus, for explicit methods to realize their full potential the
previous partial equilibrium examples must be extended to include very stiff
networks containing hundreds (or more) species. That work is ongoing and will be
reported in future publications, but we note that recent encouraging results in
this direction are discussed in Ref.\ \cite{guidPE}.

\section{\label{sh:summary} Summary}

Previous discussions of integrating extremely stiff reaction networks  have
usually concluded that such systems can be integrated only by implicit schemes,
because explicit schemes are unstable for timesteps large enough to be
efficient. Nevertheless, an alternative to implicit integration for stiff
systems is to modify the equations to be integrated using approximate algebraic
solutions to reduce the stiffness, and then to integrate the resulting equations
numerically by explicit means. For example, explicit asymptotic and QSS methods 
have had some success in integrating moderately stiff systems in various
chemical kinetics problems \cite{oran05,mott00,mott99}. However, these same
methods have been found wanting for the extremely stiff systems characteristic
of astrophysical thermonuclear networks, both because they failed to give
reliable results, and because the integration timestep was found not to be
competitive with implicit methods, even if the results had been correct
\cite{oran05,mott99}. We have presented evidence strongly challenging all of
this conventional wisdom.

The key to realizing this new view of explicit integration for stiff systems is
the understanding that in reaction networks there are (at least) three
fundamental sources of stiffness:

\begin{enumerate}

 \item 
{\em Negative populations}, which can evolve from an initially small positive
population if an explicit timestep is too large.

\item
{\em Macroscopic equilibration}, where the right sides of the differential
equations  $dY = F = \Fplus{} - \Fminus{}$ in Eq.~(\ref{equilDecomposition})
approach a constant.

\item
{\em Microscopic equilibration}, where the net flux in specific forward-reverse
reaction pairs $(f^+_i - f^-_i)$ on the right side of
Eq.~(\ref{equilDecomposition}) tends to zero. 

\end{enumerate}
These distinctions are crucial because the algebra required to stabilize the
explicit solution depends on which forms of stiffness are present in a network.
The first two sources of stiffness are handled well by asymptotic and QSS
methods, but microscopic equilibration can be handled efficiently using
explicit methods only by employing partial equilibrium approximations. 

By comparing the timesteps required to integrate problems using explicit and
implicit methods, and by using an implicit backward-Euler code to estimate the
relative speed for computing a timestep by explicit and implicit means in
networks of various sizes, we have shown that algebraically-stabilized explicit
integration can give correct results and competitive timesteps for even the
stiffest of networks.

\begin{enumerate}
 \item 
For smaller networks without significant microscopic equilibration, we find
systematic evidence that the explicit methods developed here can give
integration speeds comparable to that for implicit integration.

\item
For some cases without much equilibration  we find evidence that explicit
methods might be capable of outperforming implicit ones. We have estimated that
for equivalently-optimized codes the QSS method used here could be faster than
the implicit method used here by factors of $\sim 10-15$ for the nova simulation
of \S\ref{novaExplosions},  $\sim 10-15$ for the tidal supernova alpha network
of \S\ref{tidalSupernova}, and  $\sim 5$ for the tidal supernova 365-isotope
network of \S\ref{tidalSupernova} (though we have noted that these estimates are
uncertain by at least an order of magnitude).

\item 
For those cases where equilibration is significant the asymptotic and QSS
methods give correct results, but when used alone are much too slow to compete
with implicit methods.  However, we presented evidence in \S\ref{sh:testsAlpha}
and \S\ref{sh:extendFull} that if explicit partial equilibrium methods are used
to remove the stiffness associated with the approach to equilibrium, the
explicit methods again exhibit speeds that are comparable to or greater than
those for implicit methods.

\end{enumerate}
In addition, we have argued in \S\ref{improveSpeed} that these estimates may
not yet represent the best speeds for the new explicit methods. Because

\begin{enumerate}
 \item 
our timestepping algorithm is not yet optimal, 
\item
we have yet to investigate whether (as might be expected) QSS + partial
equilibrium can take larger timesteps than asymptotic + partial equilibrium, and

\item
an increase (algorithmically or through hardware) in the speed for computing
rates should preferentially benefit explicit methods relative to implicit ones,
\end{enumerate}
it is possible that the speed of explicit methods relative to implicit ones
might be increased substantially relative to the present results. 

Finally, the present results owe a considerable debt to previous work
\cite{oran05,mott00,mott99}. However, we find that our versions of asymptotic
and quasi-steady-state methods are much more accurate, and our versions of
partial equilibrium methods are much faster, than those of previous authors when
applied to extremely stiff astrophysical thermonuclear networks. The test cases
presented here represent a mix of quite extreme conditions, with temperature
changes as rapid as $\sim 10^{17}$ K/s, differences in fastest and slowest
network timescales as much as 10-20 orders of magnitude, and the fraction of
equilibrated reactions often changing rapidly between 0\% and 100\%. Hence, we
conjecture that these methods may be even more useful for problems where the
conditions are not quite as extreme as for astrophysical thermonuclear
environments.

\section{\label{conclusions} Conclusions}

This paper demonstrates that algebraically-stabilized explicit integration is
capable of timesteps competitive with those of implicit methods for various
extremely-stiff reaction networks. Since explicit methods can execute a timestep
faster than an implicit method in a large network, our results suggest that
algebraically-stabilized explicit algorithms may be capable of performing as
well as, or even substantially outperforming, implicit integration in a variety
of moderate to extremely stiff applications. Because of the linear scaling with
reaction network size for explicit methods, this fundamentally new view of
explicit integration for stiff equations is particularly important for
applications in fields where more realistic---and therefore larger---reaction
networks are required for physical simulations. Arguably, this means almost all
scientific and technical disciplines, since the sizes of reaction networks being
used in simulations to this point have been dictated more often by what was
feasible than by what was physical.  Of particular significance is that these
new explicit methods might permit coupling of more physically-realistic reaction
kinetics to fluid dynamics simulations in a variety of disciplines.

\begin{ack}
I thank Jay Billings, Reuben Budiardja, Austin Harris, Elisha Feger, and Raph
Hix for help with some of the calculations.  Discussions with Raph Hix, Bronson
Messer, Kenny Roche, Jay Billings, Brad Meyer, Friedel Thielemann, Michael
Smith, and Tony Mezzacappa have been useful in formulating the ideas presented
here, and I thank Bronson Messer for a careful reading of the manuscript.
Research was sponsored by the Office of Nuclear Physics, U.S. Department of
Energy.
\end{ack}

\vfill

\clearpage

\setcounter{section}{0}
\appendix
\section{Appendix:  Reaction Group Classification}
\protect\label{RGclassificationApp}

\noindent
Applying the principles discussed \S\ref{reactionGroupClasses} to the reaction
group classes in Table \ref{tb:reactionGroupClasses} gives the following partial
equilibrium properties of reaction group classes for astrophysical thermonuclear
networks.

\vspace{10pt}

\leftline{{\bf Reaction Group Class A} (a $\rightleftharpoons$ b)}

\parskip = 4pt
\parindent = 0pt

Source term: $\deriv{y_a}{t} = -k\tsub f y_a + k\tsub r y_b\quad$
Constraints:  $y_a + y_b \equiv c_1 = y_a^0 + y_b^0$

Equation: $\deriv{y_a}{t} = by_a + c \quad
b = -k\tsub f \quad c = k\tsub r\quad$
Solution:  $y_a(t) = y_a^0 e^{bt} - \frac cb \left(1-e^{bt}\right)$

Equil.\ solution:  $\bar y_a = -\frac cb = \frac{k\tsub r}{k\tsub f}\quad$
Equil.\ timescale: $\tau = \frac 1b = \frac{1}{k\tsub f}$

Equil.\ tests:  $\frac{|y_i-\bar y_i|}{\bar y_i} < \epsilon_i
\ \  (i = a, b)\quad$
Equil.\ constraint: $\frac{y_a}{y_b} = \frac{k\tsub r}{k\tsub f}$

Other variables:  $y_b = c_1 - y_a $

Progress variable:  $\lambda \equiv y_a^0 - y_a \quad y_a = y_a^0 - \lambda
\quad y_b = y_b^0 + \lambda$

\vspace{15pt}
\leftline{{\bf Reaction Group Class B} (a + b $\rightleftharpoons$ c)}

Source term: $\deriv{y_a}{t} = -k\tsub f y_a y_b + k\tsub r y_c$

Constraints:  
$y_b - y_a \equiv c_1 = y_b^0 - y_a^0\quad$
$y_b + y_c \equiv c_2 = y_b^0 + y_c^0$

Equation: $\deriv{y_a}{t} = ay_a^2 +  by_a + c\quad$
 $a = -k\tsub f \quad b = -(c_1 k\tsub f + k_b)
\quad c = k\tsub r (c_2-c_1)$

Solution: \eq{2body1.4}
$\quad$
Equil.\ solution:  \eq{2body1.6}
$\quad$
Equil.\ timescale: \eq{2body1.8}

Equil.\ tests:  $\frac{|y_i-\bar y_i|}{\bar y_i} < \epsilon_i
\ \  (i = a, b, c)\quad$
Equil.\ constraint: $\frac{y_a y_b}{y_c} = \frac{k\tsub r}{k\tsub f}$

Other variables:  
$y_b = c_1 + y_a  \quad y_c = c_2-y_b $

Progress variable:  $\lambda \equiv y_a^0 - y_a \quad y_a = y_a^0 - \lambda
\quad y_b = y_b^0 - \lambda\quad$
$y_c = y_c^0 + \lambda$

\vspace{15pt}
\leftline{{\bf Reaction Group Class C} (a + b  + c $\rightleftharpoons$ d)}

Source term: $\deriv{y_a}{t} = -k\tsub f y_a y_b y_c+ k\tsub r y_d\quad$
Constraints:
$y_a - y_b \equiv c_1 = y_a^0 - y_b^0$

$\tfrac13 (y_a+y_b+y_c)+y_d \equiv c_3 = \tfrac13 (y_a^0 + y_b^0 + y_c^0) +
y_d^0\quad$
$y_a - y_c \equiv c_2 = y_a^0 - y_c^0$

Equation: $\deriv{y_a}{t} = ay_a^2 +  by_a + c$

$a = -k\tsub fy_a^0 + k\tsub f(c_1+c_2) \quad b = -(k\tsub f c_1 c_2 + k\tsub
r)\quad$
$c = (c_3+\tfrac13 c_1 + \tfrac13 c_2) k\tsub r$

Solution: \eq{2body1.4}
$\quad$
Equil.\ solution:  \eq{2body1.6}
$\quad$
Equil.\ timescale: \eq{2body1.8}

Equil.\ tests:  $\frac{|y_i-\bar y_i|}{\bar y_i} < \epsilon_i
\ \  (i = a, b, c, d)$
$\quad$
Equil.\ constraint: $\frac{y_a y_b y_c}{y_d} = \frac{k\tsub r}{k\tsub f}$

Other variables:  
$y_b = y_a -c_1  \quad y_c = y_a -c_2$
$\quad$
$ y_d = c_3 - y_a + \tfrac13 (c_1 + c_2)$

Progress variable:  $\lambda \equiv y_a^0 - y_a \quad y_a = y_a^0 - \lambda $
$\quad$
$y_b = y_b^0 - \lambda$
$\quad$
$y_c = y_c^0 - \lambda$ 

$y_d = \lambda+y_d^0$

\newpage

\leftline{{\bf Reaction Group Class D} (a + b $\rightleftharpoons$ c + d)}

Source term: $\deriv{y_a}{t} = -k\tsub f y_a y_b + k\tsub r y_c y_d\quad$
Constraints:  
$y_a - y_b \equiv c_1 = y_a^0 - y_b^0$

$y_a + y_c \equiv c_2 = y_a^0 + y_c^0$
$\quad$
$y_a + y_d \equiv c_3 = y_a^0 + y_d^0$

Equation: $\deriv{y_a}{t} = ay_a^2 +  by_a + c$
$\quad$
$a = k\tsub r -k\tsub f \quad b = -k\tsub r (c_2 + c_3)+ k\tsub f c_1
\quad c = k\tsub r c_2 c_3$

Solution: \eq{2body1.4}
$\quad$
Equil.\ solution:  \eq{2body1.6}
$\quad$
Equil.\ timescale: \eq{2body1.8}

Equil.\ tests:  $\frac{|y_i-\bar y_i|}{\bar y_i} < \epsilon_i
\ \  (i = a, b, c, d)$
$\quad$
Equil.\ constraint: $\frac{y_a y_b}{y_c y_d} = \frac{k\tsub r}{k\tsub f}$

Other variables:  
$y_b = y_a - c_1 \quad y_c = c_2-y_a \quad y_d = c_3 -y_a$

Progress variable:  $\lambda \equiv y_a^0 - y_a \quad y_a = y_a^0 - \lambda
\quad y_b = y_b^0 - \lambda\quad$
$y_c = y_c^0 + \lambda \quad$
$y_d = y_d^0 + \lambda$

\vspace{15pt}
\leftline{{\bf Reaction Group Class E} (a + b $\rightleftharpoons$ c + d + e)}

Source term: $\deriv{y_a}{t} = -k\tsub f y_a y_b + k\tsub r y_c y_d y_e$

Constraints:  
$y_a + \tfrac13 (y_c+y_d+y_e) \equiv c_1 = y_a^0 + \tfrac13(y_c^0 + y_d^0 +
y_e^0)$

$y_a-y_b \equiv c_2 = y_a^0 - y_b^0 \quad y_c-y_d \equiv c_3 = y_c^0 - y_d^0$
$\quad$
$y_c - y_e \equiv c_4 = y_c^0 - y_e^0$

Equation: $\deriv{y_a}{t} = ay_a^2 +  by_a + c$

$a = (3c_1 - y_a^0)k\tsub r - k\tsub f \quad b = c_2 k\tsub f - (\alpha\beta
+ \alpha\gamma   + \beta\gamma) k\tsub r$
$\quad$
$ c = k\tsub r \alpha\beta\gamma$

$\alpha \equiv c_1 + \tfrac13(c_3+c_4)
\quad \beta\equiv c_1 - \tfrac23 c_3 + \tfrac13 c_4$
$\quad$
$\gamma \equiv c_1 + \tfrac13 c_3 - \tfrac23 c_4$

Solution: \eq{2body1.4}
$\quad$
Equil.\ solution:  \eq{2body1.6}
$\quad$
Equil.\ timescale: \eq{2body1.8}

Equil.\ tests:  $\frac{|y_i-\bar y_i|}{\bar y_i} < \epsilon_i
\ \  (i = a, b, c, d, e)$
$\quad$
Equil.\ constraint: $\frac{y_a y_b}{y_c y_d y_e} = \frac{k\tsub r}{k\tsub
f}$

Other variables:  
$y_b = y_a - c_2 \quad y_c = \alpha - y_a \quad y_d = \beta - y_a$
$\quad$
$y_e = \gamma-y_a$

Progress variable:  $\lambda \equiv y_a^0 - y_a \quad y_a = y_a^0 - \lambda
\quad y_b = y_b^0 - \lambda\quad$
$y_c = y_c^0 + \lambda$

$y_d = y_d^0 + \lambda \quad y_e = y_e^0 + \lambda$

\vspace{10pt}

In the equilibrium test condition we have allowed the possibility of a
different $\epsilon_i$ for each species $i$ but in practice one would 
often choose the same small value $\epsilon$ for all $i$. The results
presented in this paper have used $\epsilon_i = 0.01$ for all species.

\parindent=4ex
\parskip=0pt

\newpage

\bibliographystyle{unsrt}

\begin{thebibliography} {99}

\bibitem{oran05} E.S. Oran and J.P. Boris, Numerical Simulation of Reactive
Flow, Cambridge University Press, 2005.

\bibitem{magick}R.S. Arvidson, F.T. Mackenzie, and M. Guidry, MAGic: A
Phanerozoic Model for the Geochemical Cycling of Major Rock-Forming Components,
American Journal of Science 306 (2006) 135-190.

\bibitem{hix05}W.R. Hix and B.S. Meyer, Thermonuclear kinetics in astrophysics,
Nuc.\ Phys.\ A 777 (2006) 188-207.

\bibitem{timmes}F.X. Timmes, Integration of Nuclear Reaction Networks for
Stellar Hydrodynamics, ApJS 124 (1999) 241-263.

\bibitem{gear71}C.W. Gear, Numerical Initial Value Problems in Ordinary
Differential Equations, Prentice Hall, 1971.

\bibitem{lamb91}J.D. Lambert, Numerical Methods for Ordinary Differential
Equations, Wiley, 1991.

\bibitem{press92}W.H. Press, S.A. Teukolsky, W.T. Vettering, and B.P. Flannery,
Numerical Recipes in Fortran, Cambridge University Press, 1992.

\bibitem{fry00} B. Fryxell, K. Olson, P. Ricker, F.X. Timmes, M. Zingale, D.Q.
Lamb, P. MacNeice, R. Rosner, J. Truran, and H. Tufo, FLASH: An Adaptive Mesh
Hydrodynamics Code for Modeling Astrophysical Thermonuclear Flashes, ApJS 131
(2000) 273-334.

\bibitem{guidAsy} M.W. Guidry, R. Budiardja, E. Feger, J.J. Billings,
W.R. Hix, O.E.B. Messer, K.J. Roche, E. McMahon, and M. He, Explicit integration
of extremely-stiff reaction networks: asymptotic methods. <arXiv:1112.4716>.

\bibitem{youn77} T.R. Young and J.P. Boris, A numerical technique for solving
ordinary differential equations associated with the chemical kinetics of
reactive flow problems, J. Phys.\ Chem.\ 81 (1977) 2424-2427.

\bibitem{mott00} D.R. Mott, E.S. Oran, and B. van Leer, Differential Equations
of Reaction Kinetics, J. Comp.\ Phys.\ 164  (2000) 407-428.

\bibitem{mott99} D.R. Mott, New Quasi-Steady-State and Partial-Equilibrium
Methods for Integrating Chemically Reacting Systems, doctoral thesis, University
of Michigan, 1999.

\bibitem{verw94}J.G. Verwer and M. van Loon, An Evaluation of Explicit
Pseudo-Steady-State Approximation Schemes for Stiff ODE Systems from Chemical
Kinetics, J. Comp.\ Phys.\ 113 (1994) 347-352.

\bibitem{verw95}J.G. Verwer and D. Simpson, Explicit Methods for Stiff ODEs from
Atmospheric Chemistry, App.\ Numerical Mathematics, 18 (1995) 413-430.

\bibitem{jay97}L.O. Jay, A. Sandu, A. Porta, and G.R. Carmichael, Improved
quasi-steady-state-approximation methods for atmospheric chemistry integration,
SIAM Journal of Scientific Computing 18 (1997) 182-202.

\bibitem{guidQSS} M.W. Guidry and J.A. Harris, Explicit integration of
extremely-stiff reaction networks: quasi-steady-state methods.
<arXiv:1112.4750>.

\bibitem{guidPE}M.W. Guidry, J.J. Billings, and W.R. Hix, Explicit integration
of extremely-stiff reaction networks: partial equilibrium methods.
<arXiv:1112.4738>.

\bibitem{feg11b} E. Feger, Evaluating Explicit Methods for Solving Astrophysical
Nuclear Reaction Networks, doctoral thesis, University of Tennessee, 2011.

\bibitem{raus2000}T. Rauscher and F.-K. Thielemann, Astrophysical Reaction Rates
From Statistical Model Calculations, At.\ Data Nuclear Data Tables 75 (2000)
1-351.

\bibitem{raphcode}W.R. Hix and F.-K. Thielemann, Computational methods for
nucleosynthesis and nuclear energy generation, J. Comp.\ Appl.\ Math.\ 109
(1999) 321-351.

\bibitem{lapack}Ed Anderson, LAPACK Users' Guide, 3rd Edition, SIAM 94, 1999.

\bibitem{ma28}Iain S. Duff, A.M. Erisman and J.K. Reid, Direct Methods for
Sparse Matrices, Oxford University Press, 1986.

\bibitem{pardiso}Parallel Sparse Direct Linear Solver (PARDISO) User Guide,
Version 3.2, Computer Science Department,  University of Basel, Switzerland
(undated); O. Schenk, K. Gartner and W. Fichtner, Efficient Sparse LU
Factorization with Left-right Looking Strategy on Shared Memory Multiprocessors,
BIT 40 (2000) 158-176; O. Schenk and K. Gartner, Solving Unsymmetric Sparse
Systems of Linear Equations with PARDISO, Journal of Future Generation Computer
Systems 20 (2004) 475-487; O. Schenk and K. Gartner, On Fast Factorization
Pivoting Methods for Sparse Symmetric Indefinite Systems, Electronic
Transactions on Numerical Analysis 23 (2006) 158-179.

\bibitem{parete-koon03} S. Parete-Koon, W.R. Hix, M.S. Smith, S. Starrfield,
D.W. Bardayan, M.W. Guidry and A. Mezzacappa, Impact of a new
$\isotope{17}{F}(p,\gamma)$ reaction rate on nova nucleosynthesis, Astrophysical
Journal 598 (2003) 1239-1245.

\bibitem{novaProfile}S. Parete-Koon, Reaction Rate of $\isotope{17}{F}(p,\gamma
)\isotope{18}{Ne}$ and its Implications for Nova Nucleosynthesis, Master's
thesis, University of Tennessee, 2001.

\bibitem{feg11a}E. Feger, M.W. Guidry, and W.R. Hix, Evaluating integration
methods for astrophysical nuclear reaction networks, in preparation.

\bibitem{tidalSupernova}S. Rosswog, E. Ramirez-Ruiz, and W.R. Hix, Atypical
thermonuclear supernovae from tidally crushed white dwarfs, Astrophysical
Journal 679 (2008) 1385-1389.

\bibitem{JINA}Tabulated at http://groups.nscl.msu.edu/jina/reaclib/db/. The JINA
extensions to REACLIB are discussed in R.H. Cyburt, et al., The JINA REACLIB
Database:Its Recent Updates and Impact on Type-I X-ray Bursts, Ap.\ J.\ Supp.\
189 (2010) 240-252.

\bibitem{YASS}A. Khoklov, Yet Another Stiff Solver (YASS); result quoted in
Ref.\ \cite{mott99} and algorithm described in Ref.\ \cite{oran05}.

\end{thebibliography}

\end{document}